\documentclass[]{interact}

\usepackage{epstopdf}
\usepackage{subfigure}

\usepackage{natbib}
\bibpunct[, ]{(}{)}{;}{a}{}{,}

\theoremstyle{plain}

\theoremstyle{definition}

\theoremstyle{remark}

\usepackage{bm}
\usepackage{bbm}
\usepackage{multirow}
\usepackage{arydshln}

\newcommand{\R}{\mathbb{R}}
\newcommand{\N}{\mathbb{N}}
\newcommand{\Q}{\mathbb{Q}}
\newcommand{\e}{\mathtt{e}}
\renewcommand{\d}{\mathtt{d}}
\newcommand{\norm}[1]{\left\lVert#1\right\rVert}
\begin{document}

\articletype{FULL PAPER}

\title{Kernel Estimates as General Concept for the Measuring of Pedestrian Density}

\author{
\name{Jana Vackov\' a \textsuperscript{a}\thanks{CONTACT Jana Vackov\' a Email: janca.vackova@fjfi.cvut.cz} and Marek Buk\' a\v cek\textsuperscript{a}}
\affil{\textsuperscript{a}Faculty of Nuclear Sciences and Physical Engineering, Czech Technical University in~Prague, Prague, Czech Republic}
}

\maketitle

\begin{abstract} 
The standard definition of pedestrian density produces scattered values, hence, many approaches have been developed to improve the features of the estimated density. This paper provides a review of generally applied methods and presents a general framework based on various kernels that bring desired properties of density estimates (e.g., continuity) and incorporate ordinarily used methods. The developed kernel concept considers each pedestrian as a source of density distribution, parametrized by the kernel type (e.g., Gauss, cone) and kernel size. The quantitative parametric study performed on experimental data illustrates that parametrization brings desired features, for instance, a conic kernel with a base radius in $(0.7, 1.2)$ m produces smooth values that retain trend features. The correspondence between kernel and non-kernel methods (namely Voronoi diagram and customized inverse distance to the nearest pedestrian) is achievable for a wide range of kernel parameter. Thereby the generality of the concept is supported.
\end{abstract}

\begin{keywords} 
Pedestrian dynamics; Density; Kernel functions; Minimum distance density; Voronoi diagram
\end{keywords}

\section{Introduction}
Flow, velocity, and density are considered to be fundamental quantities in both traffic flow, e.g. \cite{treiber2013traffic}, and pedestrian dynamics, e.g. \cite{daamen2007flow, schadschneider2018pedestrian}. Although the first models discovered their essential relationship in a plain macroscopic way, e.g. \cite{underwood1960speed}, and their first improved evaluation dates back to the 1960s, e.g. \cite{edie1963discussion}, the optimal estimate of pedestrian density remains a contemporary topic of discussion.

The standard approach using physical definitions of flow and density as counting the number of pedestrians $N$ passing through a cross-section within the time interval $\Delta T$ or standing in the area $A$ enriched by the hydrodynamic approximation
\begin{equation} \label{eq:standardJrho}
J = \frac{N}{\Delta T} \quad [\textup{ped}\cdot\textup{s}^{-1}], 
\qquad \rho = \frac{N}{|A|} \quad [\textup{ped}\cdot\textup{m}^{-2}],
\qquad J = \rho \cdot v
\end{equation}
has been widely studied throughout the decades.

However, while the fundamental definitions describe the current phase of the system, the numbers themselves do not carry any specific information about the dynamics inside the system. It is possible to observe two configurations under very different internal conditions, both achieving the same value of (global) flow or (global) density, see \cite{kerner2004three}. Thus, local estimates are necessary to describe the situation in detail. Furthermore, sophisticated data analysis requires smoother data in time and space, and simple counting of cars or pedestrians results in scattered values with discrete jumps, especially for small areas.

Therefore, researchers have started to use approaches based on statistical methods (including kernel estimates) and even methods where the expression of density is based on the time to collision or the minimum distance to another car or pedestrian. Focusing on distance has become more important under pandemic conditions when guidelines require certain minimum space headways. More general methods interpreting pedestrian comfort or crowdedness as independent variables related to density have become popular due to their greater relevance for several use cases. 

\subsection{State of the Art}
The development of density estimates is discussed as follows. 

Grid-based methods by \cite{fruin1970designing} (introducing discrete density maps) localize the density contribution of each pedestrian only to the cell they are standing in at a given time. \cite{edie1963discussion} introduced a smoothed version of grid-based methods using complete trajectory information. The pedestrian's contribution to a given cell in a particular time interval is determined by the amount of time spent there. If a pedestrian stays in a grid element $A \subset \R^2$ for the entire time step $T \in \R^+$, the density contribution is $1/|A|$. Otherwise, their contribution is calculated as $(t/T) \cdot 1/|A|$, where $t < T$ is the time spent in the cell $A$ during the time interval $T$.

The smooth number of pedestrians in space and time was used in \cite{helbing2006analytical} and extended by the smooth contribution of one pedestrian in \cite{helbing2007dynamics}, which was later applied in \cite{johansson2008crowd, johansson2009data}. This definition is also an initial step towards the individual density concept with a dynamic detector, as described in \cite{bukavcek2019evaluation}. A similar effect is achieved through the weighting of pedestrians, as mentioned in \cite{schadschneider2010stochastic, schadschneider2018pedestrian}, and using the density estimate defined as the body projection into the observed area in \cite{predtetschenski1971personenstrome}, or by an inverse space headway.

The paper by \cite{steffen2010methods} about measuring fundamental quantities introduced the concept used in this paper. The authors discuss the properties of the standard definition (\ref{eq:standardJrho}) in detail and introduce the kernel approach. They consider several parametrized kernel types (Gaussian, conic, and cylindrical) together with the approach of the Voronoi diagram to estimate the density distribution in the observed area. The Voronoi method is further studied and stabilized by averaging over time in \cite{liddle2011microscopic}.

The comparison of different methods focusing on the fundamental diagram is presented in \cite{duives2015quantification}, covering weighted distance, estimates based on the minimum distance to pedestrians, or methods incorporating time spend in the detector. Another comparison of different methods, which are similar to the methods mentioned above, can also be found in \cite{hillebrand2020comparing} with the aim of measuring the safety of pedestrians.

The authors of the previously mentioned papers have pointed to the disadvantage of using the kernel approach, which requires the calibration of the scale parameter, in contrast to the popular non-parametric Voronoi approach. The paper by \cite{silverman1986density} proposed a statistical solution of this issue by utilizing the variance and interquartile range of the observations. However, this solution is not efficient since the kernel parameter needs to be recomputed for each time step. Moreover, it is questionable whether the variance of pedestrian locations should control the kernel size in the density estimation of pedestrians. Despite this, several researchers have studied this variance approach, including \cite{krisp2009visual, plaue2012multi, plaue2014measuring, mollier2019two, fan2013comparative, fan2013data}. 

It should be noted that the parametrization of any method may vary depending on its application, and the selected method can significantly affect the results. For this reason, it is not possible to determine a specific approach as the only correct one, as already discussed by researchers such as \cite{schadschneider2011empirical, hillebrand2020comparing, tordeux2015quantitative, steffen2010methods, duives2015quantification}. Furthermore, the effect of the kernel parametrization on the results has not been fully explored, and therefore there are no general instructions for it yet.

\subsection{Goals of the Paper}
The aim of this paper is to demonstrate that many applied methods share common features, despite their different methodologies. To achieve this, we present a framework that rewrites density using the kernel concept, which enables the identification of common features among various density estimates or highlights their differences. By unifying the parametrization of different methods under the general kernel concept, any further research can specify the given method using the same terminology.

Although some methods cannot be precisely expressed by kernel equations, they can still produce results that are almost identical to kernel densities with appropriate parametrization. Therefore, it can be stated that the kernel approach is a general concept that can incorporate existing methods as special cases and approach the results of non-kernel methods.

To provide the aforementioned comparison and illustrate the strengths of the kernel concept, we will evaluate the properties of density estimates on experimental data from \cite{bukavcek2015experimental}. In addition to a simple comparison and parametrization of kernels, this study will focus on density properties such as smoothness and the ability to conserve peaks, in contrast to the weaknesses of the standard definition (\ref{eq:standardJrho}). 

This paper is organized as follows. Firstly, we will define the kernel concept and present two non-kernel densities. Next, we will introduce several criteria for quantifying density estimate properties, which will allow the reader to justify the application of analysed methods more precisely. We will then assess and compare the studied methods in detail using experimental data. However, it should be emphasized that the objective of this paper is not to determine the 'best' method for evaluating density. Instead, we aim to identify differences among various kernels, discover their general features, and compare them to chosen non-kernel methods to support the universality of the kernel concept.

\section{Concept and Definition} \label{sec:def_density}
Rewrite the definition of \emph{density} in an area $A \subset \R^2$ using a distribution inspired by kernel distribution theory (studied, for instance, in \cite{wand1994kernel, baszczynska2016kernel, chacon2018multivariate})
\begin{equation} \label{eq:kernel_rho_def}
\rho = \frac{N}{|A|} = \frac{\int_A p(\bm{x}) \,\mathrm{d} \bm{x}}{|A|} 
= \frac{\int_A \sum_{\alpha=1}^N  p_\alpha(\bm{x}) \,\mathrm{d} \bm{x}}{|A|} 
= \sum_{\alpha=1}^N\frac{\int_A p_\alpha(\bm{x}) \,\mathrm{d} \bm{x}}{|A|},
\end{equation}
where $N \in \N_0$ represents the total (discrete) number of pedestrians, $|A|$ is the size of considered area $A$, $p_{\alpha}(\bm{x})$ denotes the \emph{individual density distribution} generated by each pedestrian $\alpha \in \{ 1, 2, \dots, N \}$ and 
\begin{equation}
p(\bm{x}) = \sum_{\alpha=1}^N  p_\alpha(\bm{x})
\end{equation}
defines the \emph{density distribution} in the area $A$. 

For a specific purpose, it can be useful to work with pedestrian count $C \in \R^+_0$ in the considered area $A$ instead of the density
\begin{equation} \label{eq:kernel_ped_count_def}
C = \int_{A} p(\bm{x}) \,\mathrm{d} \bm{x} = \int_{A} \sum_{\alpha=1}^N  p_\alpha(\bm{x}) \,\mathrm{d} \bm{x}.
\end{equation}
Thus, $C$ can be understood as a real alternative to a discrete number of pedestrians~$N$.

If the whole examined area is denoted as $A_0 \subset \R^2$, the individual density distribution holds the normalization condition 
\begin{equation} \label{eq:normalization_condition}
\int_{A_0} p_{\alpha}(\bm{x}) \,\mathrm{d} \bm{x} = 1. 
\end{equation}
Therefore, the relation $\int_{A_0} p(\bm{x}) \,\mathrm{d} \bm{x} = N$ is fulfilled, and the density $\rho_{A_0}$ is called \emph{global}. Moreover, for any other (sub)area $A \subseteq A_0$, pedestrian count follows $C \leq N$.

To emphasize the generality of relation (\ref{eq:kernel_rho_def}), we can write the definition of the density including kernel parameters
\begin{equation} \label{eq:density_static}
\rho_A = \frac{\int_A \sum_{\alpha=1}^N  p_\alpha(\bm{x}, R) \,\mathrm{d} \bm{x}}{|A|}.
\end{equation}
Regardless of the kernel type, $R$ expresses the smoothing factor. Thus, we refer to it as a \emph{blur}.

Blur $R$ determines the size of the area affected by an individual pedestrian. Denote this area as the pedestrian \emph{support} $A_\alpha \subset A$ and define it as follows
\begin{equation} \label{eq:support}
A_\alpha = \lbrace \bm{x} \in A \,|\, p_\alpha (\bm{x}) > 0 \rbrace ,
\end{equation}
i.e., $A_\alpha$ is the smallest possible subset of $A$ that satisfies $\int_{A_\alpha} p_{\alpha}(\bm{x}) \,\mathrm{d} \bm{x} = 1$.

When area $A \subset A_0$ covers only a part of the whole examined area $A_0$, i.e., $A \cap A_0 \not= \emptyset$ and $A \cap A_0 \not= A_0$, the pedestrian $\alpha$ may contribute to the density $\rho_A$ only partially if $A \cap A_\alpha \not= A_{\alpha}$, i.e., $0 < \int_{A} p_{\alpha}(\bm{x}) \,\mathrm{d} \bm{x} < 1$.

If the kernels intersect the walls or obstacles, the kernels are normalized to keep the pedestrian volume in the eligible area. This involves trimming the support $A_\alpha$ and rescaling the kernel to satisfy the normalization condition (\ref{eq:normalization_condition}). Therefore, the peak of the individual density distribution will be higher than in the case of non-intersection.

To be complete, the area $A$ does not have to be static (a detector approach) as in \cite{steffen2010methods}. If the area $A$ is variable in time (dynamic detector) and focuses on the surroundings of pedestrian $\alpha$, we refer to the density as the \emph{individual} density. We have studied it for preliminary results in \cite{bukavcek2019evaluation, vackova2019aplimat}. For clarity, denote $\omega_{\alpha}$ as the surroundings of pedestrian $\alpha$. Then we note $\rho_{\omega_{\alpha}}(\bm{x})$ as the \emph{individual density of pedestrian} $\alpha$, in contrast to the \emph{static detector density} $\rho_A(\bm{x})$, which will be the main topic of this paper.

\subsection{Type of Kernels}
We limit the kernel shapes to be symmetric and independent of time, since our goal is to capture and analyse the main characteristics of the density distribution. The addition of a time factor and asymmetry may improve the density estimate, and their impact can be examined separately in future research.

In addition to the Dirac delta function as the kernel, which tends to the standard approach (\ref{eq:standardJrho}), we will also use cylindrical, conic, Gaussian, and Borsalino kernels. These kernels are defined below, using the notation $\bm{x}_{\alpha} := \bm{x}_{\alpha}(t)$ to represent the head position of pedestrian $\alpha$ at a fixed time $t \in \R_0^+$.

\subsubsection{Point Approximation} 
\emph{Dirac delta function}
\begin{equation} 
p_\alpha(\bm{x}) = \delta_{\bm{x},\bm{x}_{\alpha}}
\end{equation}
represents the point approximation of the pedestrian distribution.

\subsubsection{Stepwise Function}
The stepwise function can also be used, defined as follows
\begin{equation} 
p_\alpha(\bm{x}, R) = \mathbbm{1}_{A_{\alpha}(R)}(\bm{x}) \, \frac{1}{|A_\alpha(R)|}, 
\end{equation}
where the symbol $\mathbbm{1}_X$ represents the indicator function of set $X$, and $A_\alpha(R)$ is the support of the pedestrian $\alpha$. The shape of this area is chosen arbitrarily, we choose the \emph{cylindrical kernel}, i.e.,
\begin{equation} \label{eq:ped_support}
A_{\alpha}(R) = \left\{ \bm{x} \in \R^2 : \norm{\bm{x} - \bm{x}_{\alpha}}\leq R \right\}
\end{equation}
and
\begin{equation} \label{eq:indicator}
\mathbbm{1}_{A_{\alpha}(R)}(\bm{x}) = \Theta\left(R - \norm{\bm{x} - \bm{x}_{\alpha}}\right),
\end{equation}
where $\Theta(\cdot)$ represents the Heaviside step function.	
				
\subsubsection{Conic Kernel}
The conic kernel uses the same pedestrian support (\ref{eq:ped_support}) and indicator function (\ref{eq:indicator}) as the cylindrical kernel. However, it has a linear trend (not the constant as previously)
\begin{equation} 
p_{\alpha}(\bm{x}, R) = \frac{3}{\pi R^3} \mathbbm{1}_{A_{\alpha}(R)}(\bm{x}) \, \left(R - \norm{\bm{x} - \bm{x}_{\alpha}}\right).
\end{equation}
It has several desired features for representing pedestrians compared to the stepwise kernels, such as decreasing trend with increasing distance, limited support, and independence of one pedestrian from the others. 

\subsubsection{Borsalino Kernel} 
The Borsalino kernel comes from \cite{krbalek20183s}, where it is defined for one dimension. The generalization used in this paper for two dimensions can be defined as follows
\begin{equation} 
p_{\alpha}(\bm{x}, R) = \frac{1}{Z} \, \frac{1}{R} \, \mathbbm{1}_{A_{\alpha}(R)}(\bm{x}) \, \exp \left\{-\frac{1}{1 - \norm{\bm{x} - \bm{x}_{\alpha}}^2/R^2} \right\}
\end{equation}
with the same pedestrian support (\ref{eq:ped_support}) and indicator function (\ref{eq:indicator}) as in the previous cases and a normalization constant $Z$, which is numerically calculated. This kernel has similar properties as the conic kernel, however, they differ in theoretical features. The Borsalino kernel is differentiable, and this fact can be helpful in analytical calculations. On the contrary, the conic kernel does not have to be normalized using numerical computation. 
	
\subsubsection{Gaussian Kernel} 
We use the Gaussian kernel in its symmetric version with a diagonal covariance matrix
\begin{equation} 
p_{\alpha}(\bm{x}, R) = \frac{1}{2 \pi R^2} \, \exp\left\{ -\frac{\norm{\bm{x} - \bm{x}_{\alpha}}^2}{2 \, R^2} \right\}.
\end{equation}
This is the only representative with a limitless area of influence (i.e., with unbounded pedestrian support) in this study. The blur parameter $R$ represents the kernel bandwidth in our study, thus it is comparable through all used kernel shapes. 

\subsection{Non-Kernel Densities} \label{sec:nonkernel}
Having defined the kernel concept, we will present two non-kernel methods: the Voronoi diagram and the minimum distance estimate.

The Voronoi diagram was chosen to be compared with the kernel density concept as it is one of the frequently used methods for density estimation in pedestrian dynamics. In contrast, the minimum distance density estimate brings another perspective to pedestrian density and reflects the pedestrian experience in terms of the level of crowdedness, which has increased in importance after the COVID-19 pandemic in 2020.

It is important to note that kernel density is a product of the method presented earlier in this paper, which involves using individual density distribution as a function with a specific definition that does not change due to the conditions in the surroundings. Although the Voronoi and minimum distance methods can be understood as kernel densities with a specific kind of kernel that is influenced by the kernels (pedestrians) around them, they are not the kernel method themselves.

\subsubsection{Voronoi Diagram}
Each spatial point $\bm{x} \in \R^2$ is assigned to the nearest pedestrian $\alpha$ at position $\bm{x}_{\alpha}$. The set of these points is called the Voronoi cell in \cite{steffen2010methods}, denoted as $A_{\alpha}$ for pedestrian $\alpha$. Thus, the area $A_{\alpha}$ may be interpreted as the pedestrian support as it was in the context above. However, $A_\alpha$ does not depend on any blur parameter $R$ here.

It is possible to use a supplementary parameter as an upper boundary for the size of the Voronoi cell as discussed in \cite{steffen2010methods}. Nevertheless, we do not implement this parameter in our study. It would have a different meaning (maximum blur) than a simple blur of pedestrians used in other kernels, thus it would not have been comparable. Besides, the maximum blur would only influence the low occupied room, i.e., only the beginning of the density time series under the conditions of our experiment (see details in further section). However, it is necessary to consider using the maximum blur in general situations.
	
\begin{figure}[h!]
	\begin{center}
	\includegraphics[width=1\textwidth]{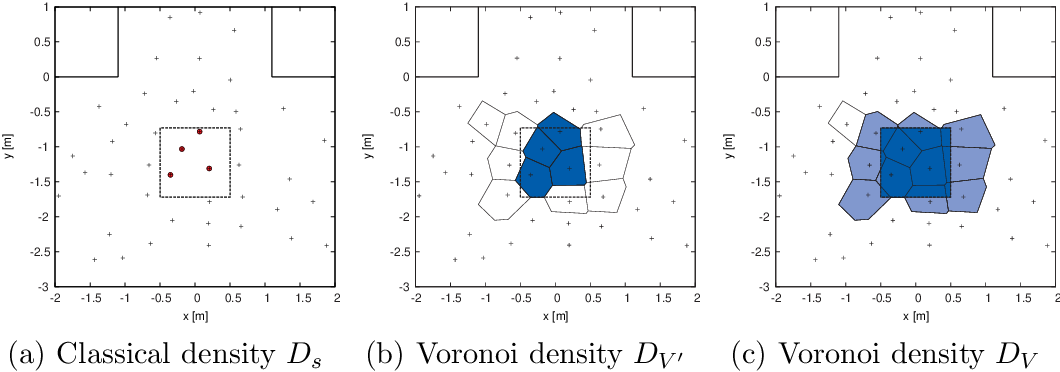}
	\caption{Application of Voronoi diagram according to \cite{steffen2010methods}. Figure taken from \cite{steffen2010methods}.} \label{fig:steffen_voronoi}
	\end{center}
\end{figure}

As introduced in \cite{steffen2010methods} and depicted in Figure \ref{fig:steffen_voronoi}, there are multiple ways to evaluate density in a detector using the Voronoi approach. It is possible to incorporate cells that have a cross-section with the detector, or only cells of pedestrians with their head positions in the detector, or to cut cells to an exact detector area. In this study, we have implemented the exact approach.	
	
\subsubsection{Minimum Distance Estimate}
Imagine a one-dimensional space, where we can write the standard definition (\ref{eq:standardJrho}) of density as $\rho = N/x$ with the number of agents $N \in \N$ and the length of the interval $x \in \R$. It can be rewritten using the mean space headway between consecutive agents as $\rho = 1/\langle x \rangle$. Thus, we can conclude that to find the maximum density, we need to find the minimum possible mean space headway. Hence, the highest density that can be measured is the density obtained by the estimate $\rho_{\textup{max}} \approx 1/\min \langle x \rangle$.

However, we operate in two dimensions in pedestrian dynamics. Then, we can rewrite the standard definition similarly as $\rho = 1/ \langle |A| \rangle$, where $\langle |A| \rangle$ can be imagined as the size of the mean area occupied or held by one pedestrian, i.e., the mean area $A_{\alpha}$ over possible pedestrians $\alpha$. For this reason, we can approximate $\langle |A| \rangle$ as a circle with a radius equal to the mean distance to the neighbour, i.e., $\langle |A| \rangle = \pi \langle x \rangle^2$. This means that the highest density is again connected with the minimum distance to a neighbour.

We define the density distribution (at every time) as 
\begin{equation}
p(\bm{x}) = \frac{c_{md}}{\textnormal{dist}(\bm{x})},
\end{equation}
where $c_{md} \in \R^+$ represents a calibration constant. We will examine two options for the function $\textnormal{dist}(\bm{x})$.

According to \cite{duives2015quantification}, we will use the minimum distance measure with a specific field of vision, defined as
\begin{equation}
\mathrm{dist}(\bm{x}):= D(\bm{x}) := \min_{\{\alpha \in \mathrm{N}: \measuredangle \left( \bm{x}_{\alpha}, \bm{s}_{\bm{x}} \right) \leq \frac{\pi}{3} \}} \norm{\bm{x} - \bm{x}_{\alpha}},
\end{equation}
where $\bm{s}_{\bm{x}}$ represents the direction from the current point $\bm{x} \in \R^2$ to the desired location, which is the middle point of the exit in this case (see the further section). In \cite{duives2015quantification}, the minimum distance density is evaluated only in the pedestrian locations $\bm{x}_{\alpha}$. 

Different from \cite{duives2015quantification}, we evaluate the density distribution at every point in space. Therefore, we have to set the field of vision at every point in space, find the pedestrian that is closest to this point (if any), and count the value of the distribution. The angle of the pedestrian field of vision is chosen according to \cite{duives2015quantification} as $120$ degrees, i.e., $2/3 \pi$ radians.

Note that assigning the field of vision to all points in space is only possible in scenarios where all pedestrians have a single global goal, as is the case in this study (see the further section). In scenarios with multiple directions, this approach would need modification, such as removing the field of vision or using a tailored approach specific to the scenario. However, research on this topic is beyond the scope of this paper.  

Due to the two-dimensional space we operate in, we are highly motivated to check our two-dimensional improvement of the previous distance measure defined as
\begin{equation}
\mathrm{dist}(\bm{x}) := D(\bm{x})^2.
\end{equation}
It is expected that this measure will bring better results because we are aware of the approximation discussed above.

However, we have to note that there are undefined points in the $\varepsilon$-neighbourhood of pedestrians ($\varepsilon \in \R^+$, fixed as $\varepsilon = 0.05$ m here) due to the avoidance of density distribution values tending to infinity, or the absence of any pedestrian in the field of vision from a specific point. These undefined points are excluded from density computation in this study. Note that the undefined points in the $\varepsilon$-neighbourhood of the minimum distance distribution can also be resolved by redefining the minimum distance measure as follows
\begin{equation}
\mathrm{dist}(\bm{x}) := \max \left\{\varepsilon, D(\bm{x})^2 \right\}.
\end{equation}
This formulation results in a shifted calibration constant $c_{md}$. Nevertheless, it does not have a significant impact on the results for low values of $\varepsilon$.

\subsection{Visual Comparison}
\begin{figure}[h!]
	\begin{center}
	\includegraphics[width=1.1\textwidth]{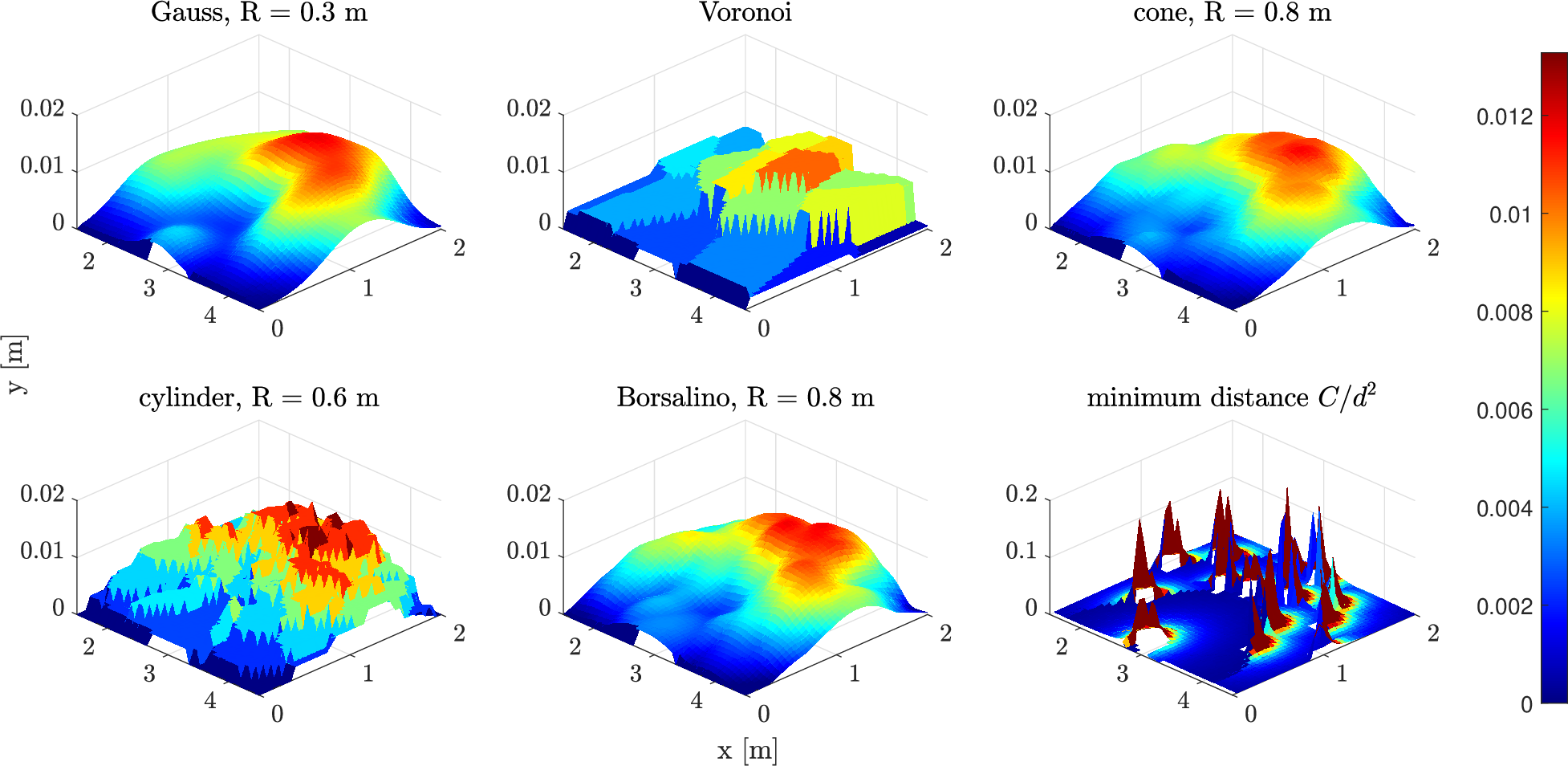}
	\caption{An example of density distributions in a sub-area (side view). The kernel size $R$ was chosen to ensure the same variance for all kernel methods, thus showing the main visual differences between different kernels (the choice of blur is studied deeply in the further text).} \label{fig:density_distr_1}
	\end{center}
\end{figure}

\begin{figure}[h!]
	\begin{center}
	\includegraphics[width=1.1\textwidth]{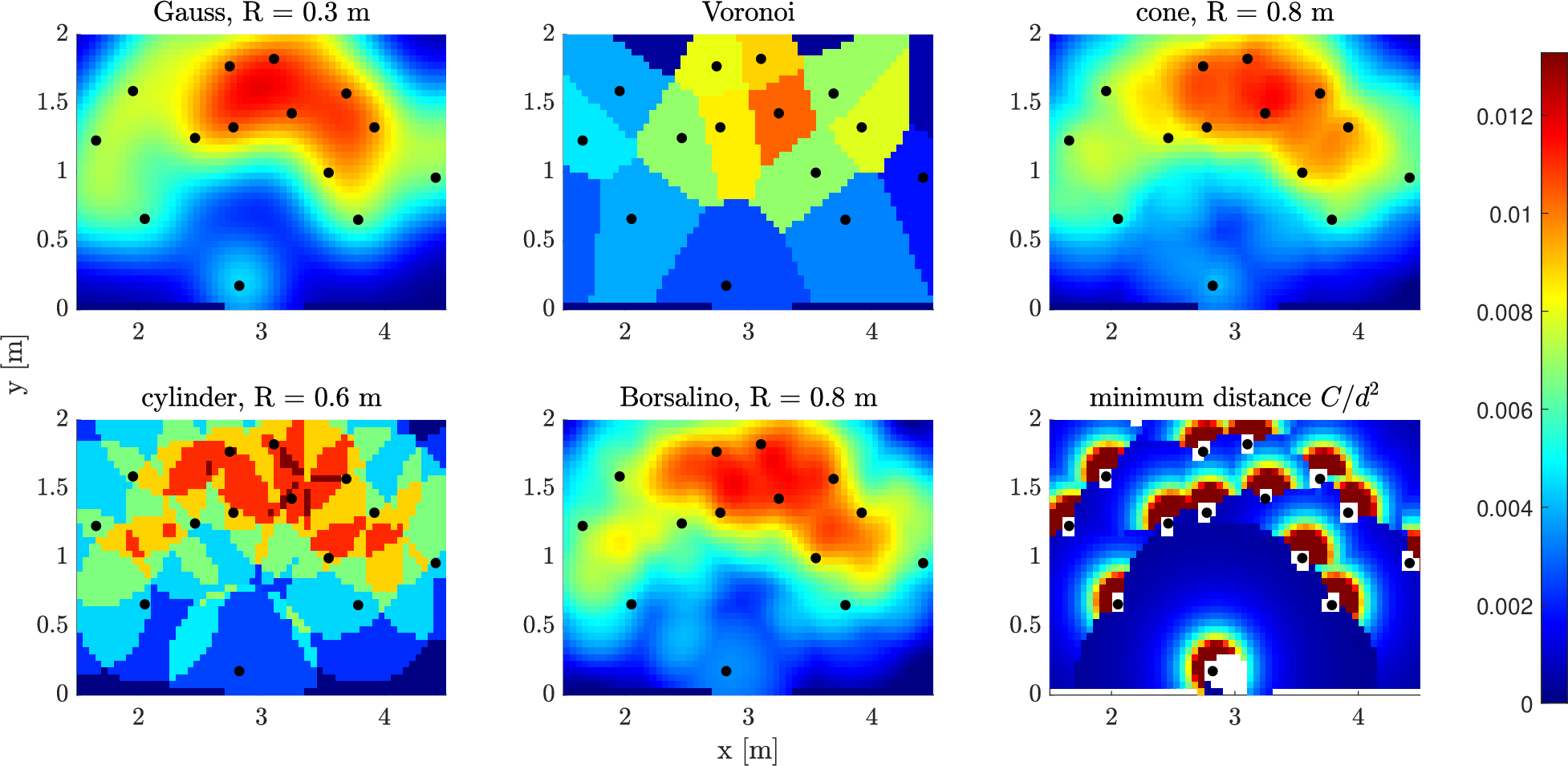} 
	\caption{An example of density distributions in a sub-area (view from above). The kernel size $R$ was chosen to ensure the same variance for all kernel methods, thus showing the main visual differences between different kernels (the choice of blur is studied deeply in the further text).} \label{fig:density_distr_2}
	\end{center}
\end{figure}

The example of density distribution for different kernels can be seen in Figure \ref{fig:density_distr_1} and Figure \ref{fig:density_distr_2}. The value of the blur $R$, which represents the kernel size, was chosen to ensure the same variance for all kernel methods. The fixed variance (using the blur value for each kernel) was selected based on metrics such as roughness and featurelessness, which will be defined and studied in the following sections, to be close to the properties of the non-kernel methods.

Although the variance of all kernels was normalized for visualization, there are significant differences in their smoothness. The Voronoi method and cylindrical kernel exhibit a jump-like behaviour, while the Gaussian, cone, and Borsalino kernels are smooth. These three kernels differ primarily in their ability to localize high-density areas, based on their limited or unlimited support. The Borsalino kernel returns several density peaks in the pedestrian overlapping area (similarly to the cylinder), while the cone and mainly the Gaussian kernel deliver one smooth maximum. The density distribution of the minimum distance estimate also differs, with undefined values observed in the area around pedestrians and the high-density area behind them. The impact of blur on density and the possibility of matching the minimum distance estimate to the kernel approach are explored in detail in the following section.

\vspace{20pt}

\newpage
\subsection{Quantitative Measures} \label{sec:methodology}
As mentioned previously, the standard definition of density (\ref{eq:standardJrho}) has several negative properties that result in volatile density time series. To quantify the volatility of all methods under consideration, we define the mean absolute derivative understood as the \emph{roughness} of the curve of density $\rho$ as follows
\begin{equation} \label{eq:def_roughness}
\langle \dot{\rho} \rangle := \frac{1}{\# \, T} \sum_{i=2}^{\# \, T} \frac{\left| \rho(t_i) - \rho(t_{i-1}) \right|}{t_i - t_{i-1}} \qquad \textup{[ped/m}^2\textup{/s]},
\end{equation}
where $\# \, T$ denotes the total number of frames captured at times $\left( t_i \right)_{i=1}^{\# T}$ available from the experiment. 

On the other hand, any smoothing should not excessively decrease the range of the curve. This property is described by the maximum the density
\begin{equation} \label{eq:def_featurelessness}
\max_{t \in \R^+} \rho(t) \qquad \textup{[ped/m}^2\textup{]},
\end{equation}
which can be understood as the level of \emph{featurelessness}.

To describe any differences on average between two densities, we use the \emph{mean absolute deviation}
\begin{equation}  \label{eq:mean_abs_diff}
\left\langle |\rho_1 - \rho_2 | \right\rangle := \frac{1}{\# \, T} \sum_{i=1}^{\# \, T} \left| \rho_1(t_i) - \rho_2(t_{i}) \right|  \qquad \textup{[ped/m}^2\textup{]}. 
\end{equation}

The mean absolute deviation is sufficient to detect similarity in the time development of two curves if it is close to zero. However, if it is significantly greater than zero, it does not enable us to identify any other relationship between the curves. Thus, we define two metrics, namely \emph{success} and \emph{integral ratio}, respectively, as follows
\begin{equation} \label{eq:success_ratio}
S_{\leq}(\rho_1, \rho_2) := \frac{1}{\# T} \sum_{i=1}^{\# T} \mathbbm{1}_{\left\{\rho_1(t_i) \leq \rho_2(t_i) \right\}} \qquad \textup{[-]}
\end{equation}
and
\begin{equation} \label{eq:integral_ratio}
I(\rho_1, \rho_2) := \frac{\int_{\R_0^+} \rho_2(t) \, \mathrm{d} t}{\int_{\R_0^+} \rho_1(t) \, \mathrm{d} t} \qquad \textup{[-]}.
\end{equation}

Despite the aforementioned drawbacks, the standard definition (\ref{eq:standardJrho}) produces solid macroscopic values. Thus, any other approach should be close to the standard approach for sufficiently macroscopic and long-term estimates. As the detector is significantly larger than the pedestrian size, any pedestrian cannot avoid it. Therefore, the mean density in a detector should be similar for all considered methods
\begin{equation} 
\label{eq:n_conserv}
\langle \rho \rangle =  \dfrac{1}{t_{1} - t_0} \int_{t_0}^{t_{1}} \rho(t) \,\mathrm{d}t = \dfrac{1}{t_{1} - t_0} \int_{t_0}^{t_{1}} \dfrac{N(t)}{A} \,\mathrm{d}t = \dfrac{\langle N \rangle}{A},
\end{equation}
where $\langle \cdot \rangle$ denotes the mean value of a quantity, $N(t) \in \N_0$ represents the total number of pedestrians in area $A$ at time $t \in \langle t_0, t_1 \rangle$, where $t_0, t_1 \in \R^+_0, t_1 > t_0$.

\section{Experimental Data-Driven Study}
The following parametric study is based on the egress experiment organized in the study hall of FNSPE, CTU in Prague in 2014, as detailed in \cite{bukavcek2015experimental, bukavcek2016individual, bukavcek2018microscopic}. Pedestrians (undergraduate students wearing recognition caps) entered the room through one of three entrances, walked to the opposite wall, and left the room through one exit. They were instructed to enter the room when the green signal was given and to leave the room as quickly as possible without running or engaging in any physical contact (to prevent any injuries).

\begin{figure}[h!]
	\begin{center}
	\hfill\includegraphics[height=.28\textwidth]{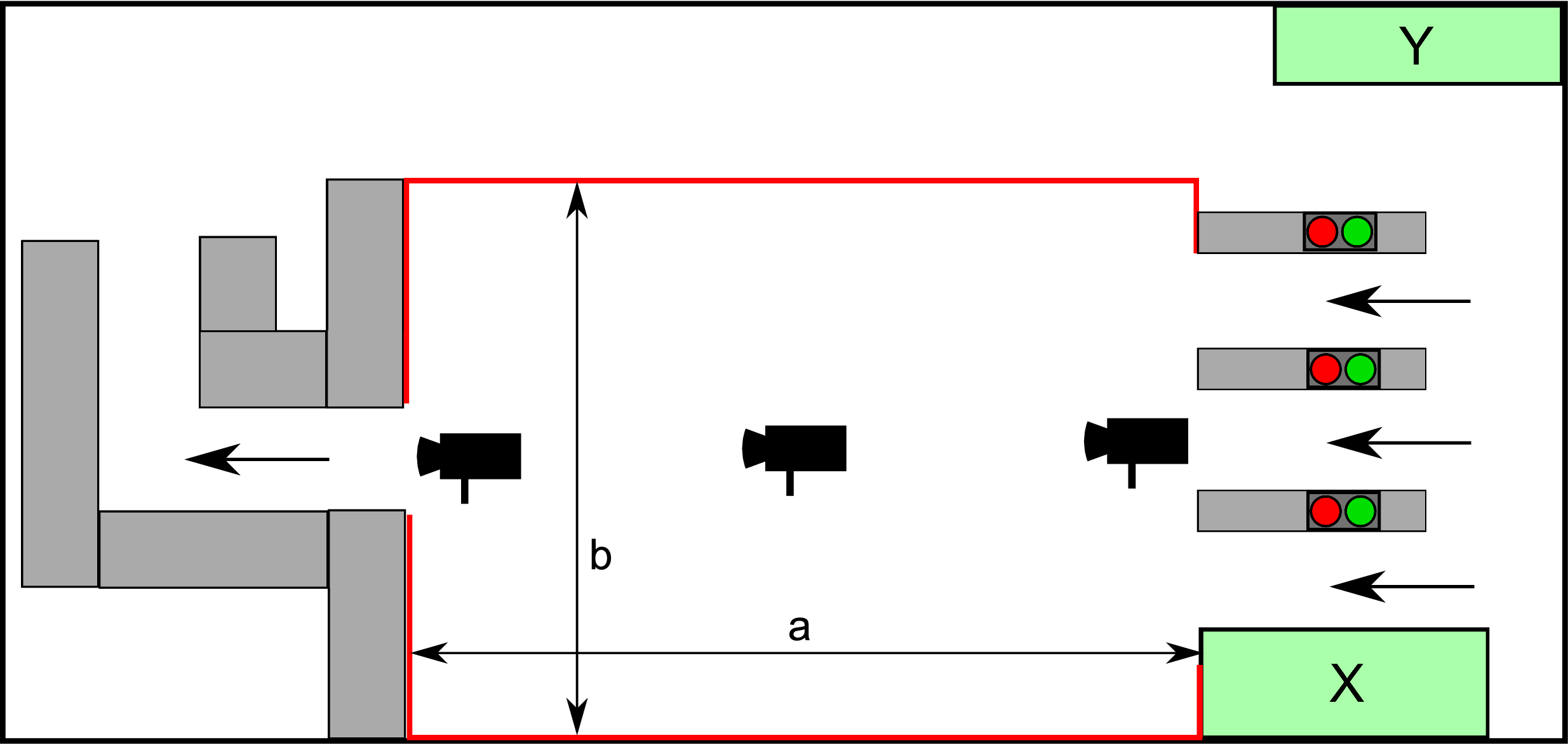}\hfill
	\includegraphics[height=.28\textwidth]{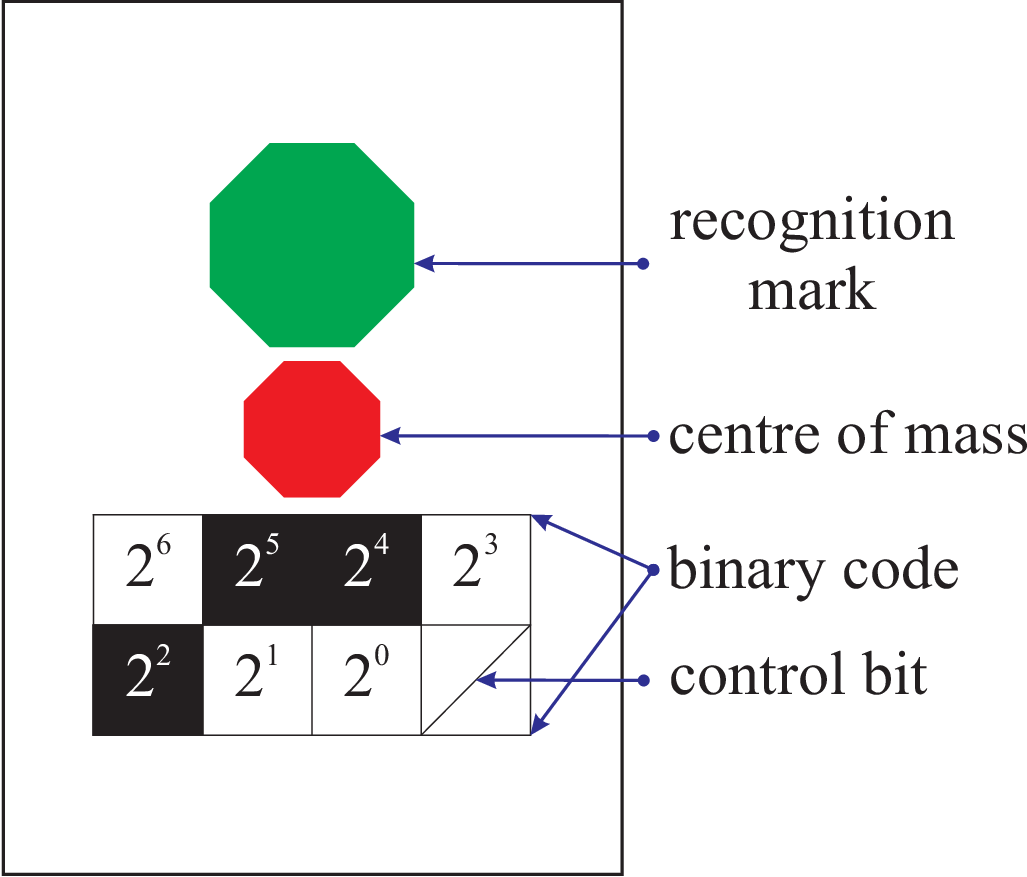}\hfill\phantom{x}
	\end{center}
\caption{Left: Experimental setting. The dimensions of the experimental room are as follows: $a = 8$ m, $b = 4.5$ m. Black symbols represent the position of cameras, technical support was situated in area X, and area Y represents the refreshment corner. Right: a sketch of a pedestrian's cap, with its size corresponding to an A4 sheet.}
\label{fig:setting}
\end{figure}

\begin{figure}[h!]
	\begin{center}
	\includegraphics[width=.99\textwidth]{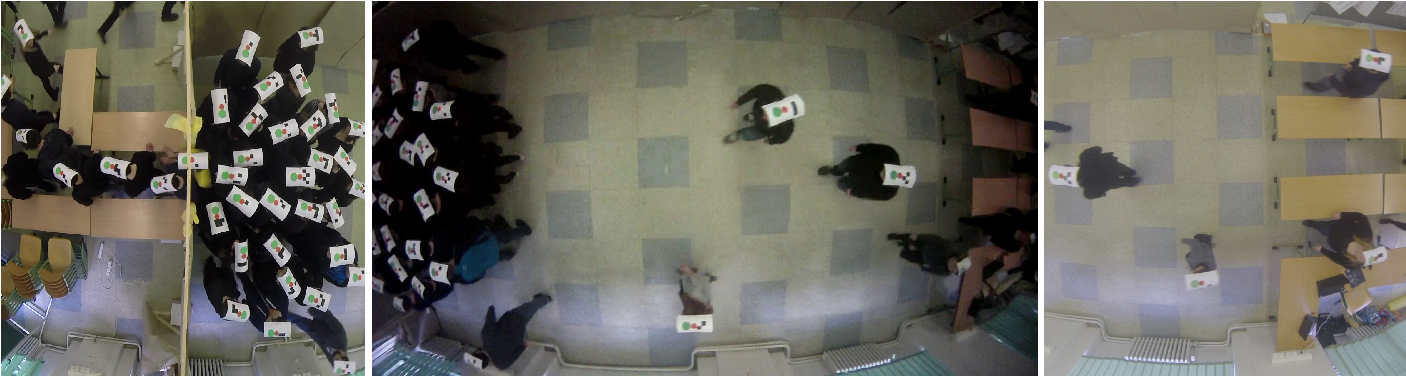}
	\end{center}
\caption{Snapshot from all three cameras at 2 min 21 s of the experiment, round number 7. A cluster of approximately $25$ pedestrians can be seen.}
\label{fig:snapshot}
\end{figure}

The observed room was equipped with three cameras: one above the center of the room, one at the entrance, and one at the exit. A snapshot from the experiment is presented in Figure~\ref{fig:snapshot}. The sample consists of 2000 paths from $10$ experimental rounds, captured using $48$ frames per second, i.e., $(t_i - t_{i-1})^{-1}$ from (\ref{eq:def_roughness}) as the frame rate is equal to $48$ s$^{-1}$. Each participant was equipped with a hat with a unique binary code enabling automated recognition, as shown in Figure~\ref{fig:setting}.

The red signal randomly generated from geometric distribution alternated with $0.1$ s green signal, allowing one person to enter. By randomized input flow control, different conditions from free flow to congestion in the exit area were achieved. Free flow was observed at the beginning of each round, as the room was always empty, and throughout the whole two rounds with a low inflow rate. On the other hand, during several rounds, the density exceeded $2$ ped/m$^2$ with local peaks over $4$ ped/m$^2$. In summary, we have in use two rounds of free flow, two rounds of a transition phase, three rounds of a stable cluster, and three rounds of congestion.

Considering the observed configurations, the data sample is sufficient for a general study. More general experimental designs, such as a junction, counterflow, or infrastructure with obstacles, would result in different values of density. However, they cannot break the presented concept itself. The evaluated density (independent of pedestrian motivation and even their velocities) depends only on the pedestrian locations produced by any design. 

\begin{figure}[h!]
	\begin{center}
	\includegraphics[width=0.55\textwidth]{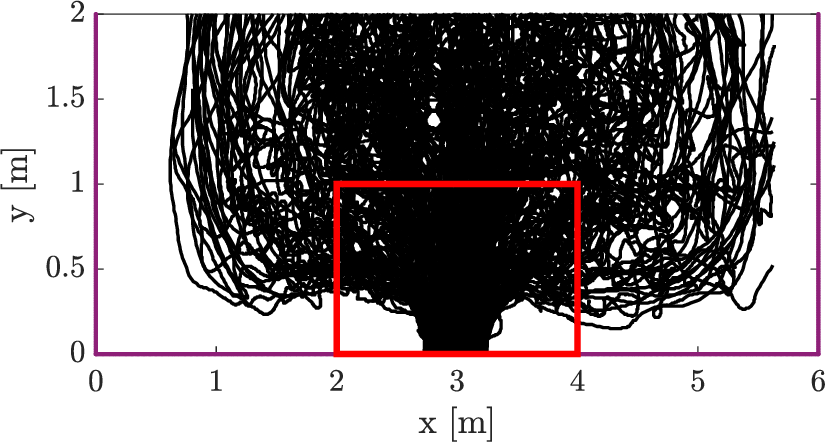}
	\caption{Trajectories in round 6 of the experiment and the chosen static detector (red rectangle).} \label{fig:rnd6_detector_traj}
	\end{center}
\end{figure}

To examine the kernel shape and size, a parametric study is based on a simple static detector visualized as the red rectangular area $A$ in Figure \ref{fig:rnd6_detector_traj}
$$
A := \left\{ \bm{x} = (x_1, x_2) \in \R^2 : x_1 \in \langle 2, 4 \rangle \, \wedge \, x_2 \in \langle 0, 1 \rangle \right\}.
$$
The actual detector size is $2.1525$ m$^2$ due to the edge pixels of the spatial grid included. The notation $\rho(t) := \rho_A(t)$ is used for simplicity in this paper.

To compare experimental rounds with different time lengths, the normalized time $t_{norm}$ will be used. The start is represented by $t_{norm} = 0$, and the end is represented by $t_{norm} = 1$.

Conic, cylindrical, Gaussian, and Borsalino kernels are evaluated for a wide range of parameter $R \in \left\{0.1, 0.2, \dots, 3 \right\}$, while Voronoi distribution, point approximation, and minimum distance approach are evaluated only once (there is no parameter). 

In our study, firstly, we will firstly verify the possibility of conic kernel copying all previously mentioned density evaluation methods. Then the quantitative comparison will be provided.

\subsection{Convergence to Point Approximation} \label{sec:converg_dirac}
It is known from the theory of generalized functions
\begin{equation}
\lim_{R \to 0^+} p(\bm{x}, R) = \lim_{R \to 0^+} \sum_{\alpha=1}^N p_{\alpha}(\bm{x}, R) = \sum_{\alpha=1}^N \delta_{\bm{x}, \bm{x}_{\alpha}},
\end{equation}
i.e., the sum of individual density distributions converges to the sum of centralized Dirac delta functions at the pedestrian locations.

\begin{figure}[h!]
	\begin{center}
	\includegraphics[width=0.55\textwidth]{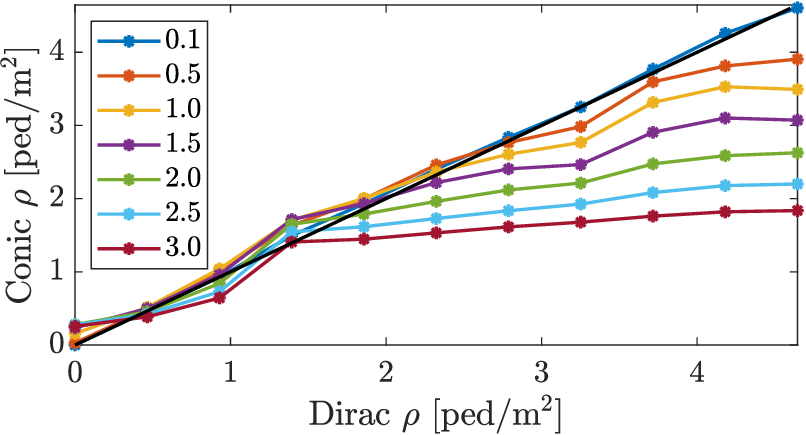}
	\end{center}
	\caption{Comparison of the density obtained by the point approximation and the median value of conic kernel density for a fixed value of Dirac density.} 
	\label{fig:simpoint}
\end{figure}

To illustrate this property, Figure \ref{fig:simpoint} was prepared to compare the density obtained by the point approximation and the density of the examined conic kernel, specifically the median value of the conic density for the fixed value of the Dirac density. It is evident that the closer the axis of the first quadrant, the closer the Dirac density is.

Figure \ref{fig:simpoint} shows that the conic density for blur $R = 0.1$ m is very similar to the the Dirac density, which is in accordance with the theoretical result. Moreover, the deviation between the conic and Dirac densities is greater for higher Dirac density and higher values of blur. Specifically, the higher the blur and the Dirac density, the lower the slope of the line. This is caused by the blending of the pedestrian outside the detector, which is more significant for a greater Dirac density.

\subsection{Voronoi Alternatives} \label{sec:voronoi_alt}
\begin{figure}[h!]
	\begin{center}
	\includegraphics[width=1\textwidth]{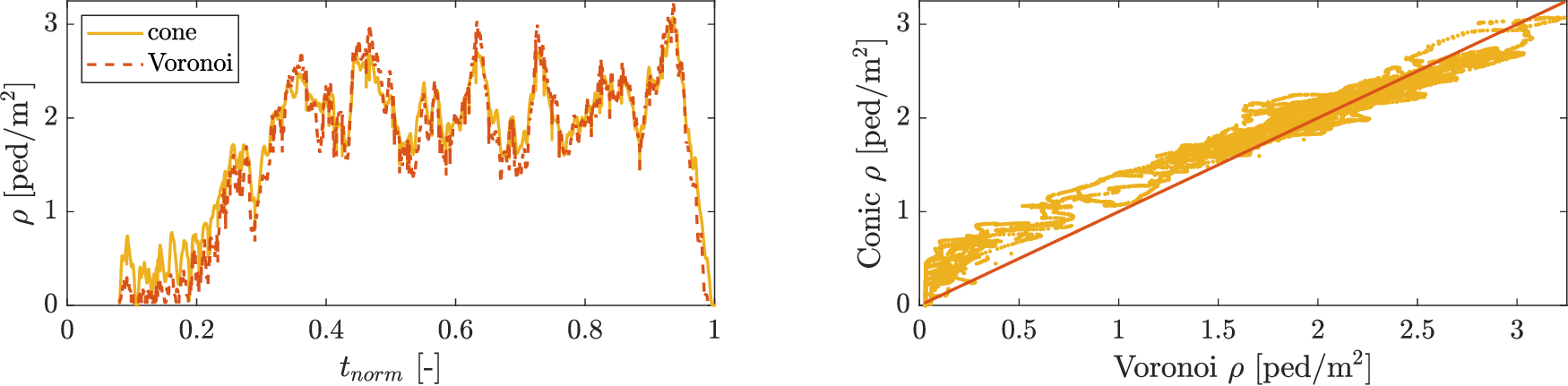}
	\end{center}
	\caption{Conic alternative with $R = 1.6$ m to Voronoi diagram for experimental round number 6: time development (left), density distribution (right).} 
	\label{fig:rnd6_alternative_voronoi}
\end{figure}

It may be useful to realize if there are any parametric settings of the conic kernel that are close to the Voronoi density. To measure this similarity, we use the mean absolute deviation (\ref{eq:mean_abs_diff}), i.e., the closer the mean absolute deviation is to zero, the more similar the densities are. An example of the time development of density (left) and density distribution for the conic kernel with $R = 1.6$ m compared to the Voronoi distribution (right) is shown in Figure \ref{fig:rnd6_alternative_voronoi}. 

Having computed the mean absolute deviation for all possible values of $R$ and experimental rounds, we can conclude that the Voronoi density has an alternative in the conic kernel, specifically for $R \in (1.5, 3)$ m. However, the value of $R$ depends on the phase of the system (experimental round). Additionally, there is a greater range for Voronoi than for the cone, and slight differences can be seen for low densities in Figure \ref{fig:rnd6_alternative_voronoi} (left). This undervaluation of the Voronoi density is evident in Figure \ref{fig:rnd6_alternative_voronoi} (right) and is caused by blending of pedestrians during free flow phases. The greater range is due to the property that the Voronoi diagram resembles the Dirac distribution under the specific condition of a dense crowd.  

\subsection{Recomputation of Gaussian Mass} \label{sec:mass_gauss}
As shown in Figure \ref{fig:density_distr_2}, the Gaussian kernel produces a very similar distribution to the conic kernel, however, more blurred due to the Gaussian limitless support. We would like to investigate if a formula exists that relates the Gaussian and conic kernels to reduce this difference.

To achieve this goal, we aim to shift almost all of the mass of the Gaussian kernel to the base of the cone centred at $\bm{0}$ denoted as $A_R = \left\{ \bm{x} \in \R^2 : \norm{\bm{x}} \leq R \right\}.$ Denote the Gaussian kernel at $\bm{0}$ with variance $\sigma^2 \in \R^+$ as
$$
p(\bm{x}, \sigma) = \frac{1}{2\pi \sigma^2} \, \e^{- \frac{\norm{\bm{x}}^2}{2 \sigma^2}}.
$$

\begin{figure}[h!]
	\begin{center}
	\includegraphics[width=0.55\textwidth]{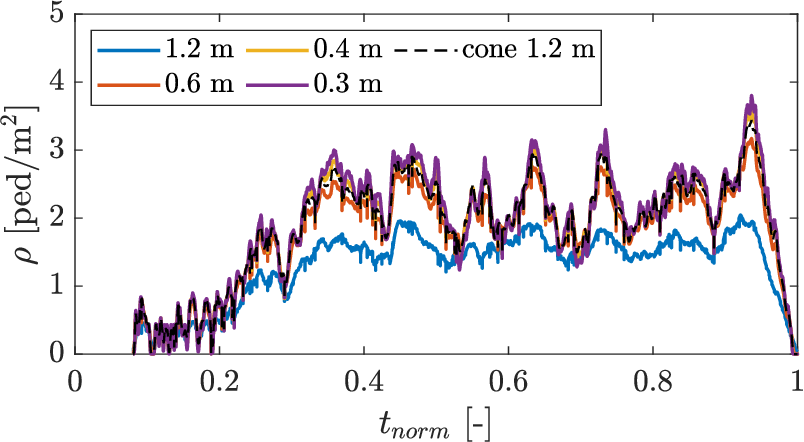}
	\end{center}
	\caption{Comparison of Gaussian and conic densities in time for experimental round number 6, using different values of $R$.} 
	\label{fig:rnd_6_Gauss_cone_simil}
\end{figure}

In order to find a relation between the conic $R$ and Gaussian $\sigma$, we assume that $R = k \sigma$, $k \in \N$. We can then solve the following integral
$$
\int_{A_R} p(\bm{x}, \sigma) \, \d \bm{x} = \int_{A_{k \sigma}} p(\bm{x}, \sigma) \, \d \bm{x} = \int_0^{\frac{k^2}{2}} \e^{-t} \d t.
$$
The result is the \emph{lower incomplete Gamma function}, which has numerically computable values. Gaussian mass inside the base of the cone for $k \in \left\{ 1, 2, 3, 4 \right\}$ is presented in Table \ref{tab:gauss_mass}. For instance, when $\sigma = \frac{R}{3}$, the mass of the Gaussian kernel inside the base of the cone with radius $R$ is approximately $99\,\%$. Thus, their densities are comparable, as illustrated in Figure \ref{fig:rnd_6_Gauss_cone_simil}.

\begin{table}[h!]
\tbl{Gaussian mass inside the base of the cone for different values of $k$.}
{\begin{tabular}{|r||c|c|c|c|}
\hline 
$k$ [-] & $1$ & $2$ & $3$ & $4$ \\ 
\hline 
mass [\%] & $39.35$ & $86.47$ & $98.89$ & $99.97$ \\ 
\hline 
\end{tabular}} \label{tab:gauss_mass}
\end{table}

\subsection{Comparison of Conic and Borsalino Kernels}
\begin{figure}[h!]
	\begin{center}
	\includegraphics[width=0.55\textwidth]{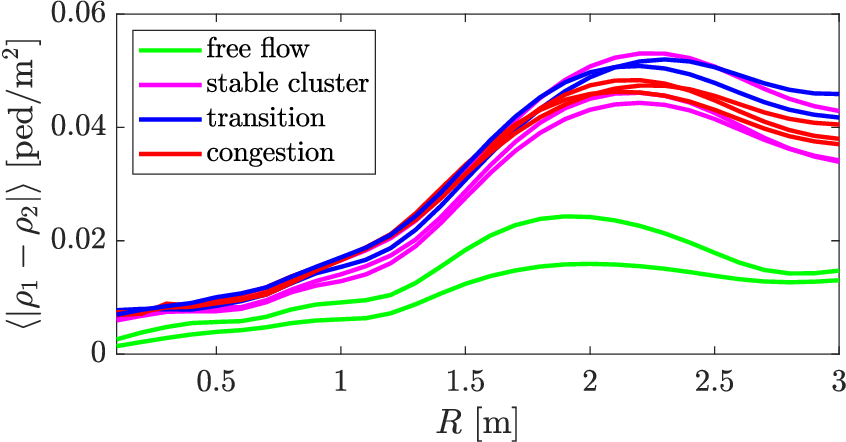}
	\end{center}
	\caption{Mean absolute deviation between the conic $\rho_1$ and Borsalino density $\rho_2$ for all experimental rounds (free flow - green, transition - blue, stable cluster - magenta, congestion - red).} \label{fig:borsalino_cone_meandifference}
\end{figure}

Considering the properties of the conic and Borsalino kernels, including their limited support and decreasing trend, we expect them to behave similarly. To evaluate this, we calculated the mean absolute deviation (\ref{eq:mean_abs_diff}) between the densities of the conic and Borsalino kernels which can be seen in Figure \ref{fig:borsalino_cone_meandifference} for different experimental rounds. 

There are dissimilarities between the free flow phase (green) and the other experimental rounds (transition - blue, stable cluster - magenta, congestion - red). Nevertheless, the mean absolute deviation is close to zero for all values of blur $R$, and the greatest difference is less than $0.06$ ped/m$^2$. Thus, the densities of the conic and Borsalino kernels coincide on average. 

\begin{figure}[h!]
	\begin{center}
	\includegraphics[width=1\textwidth]{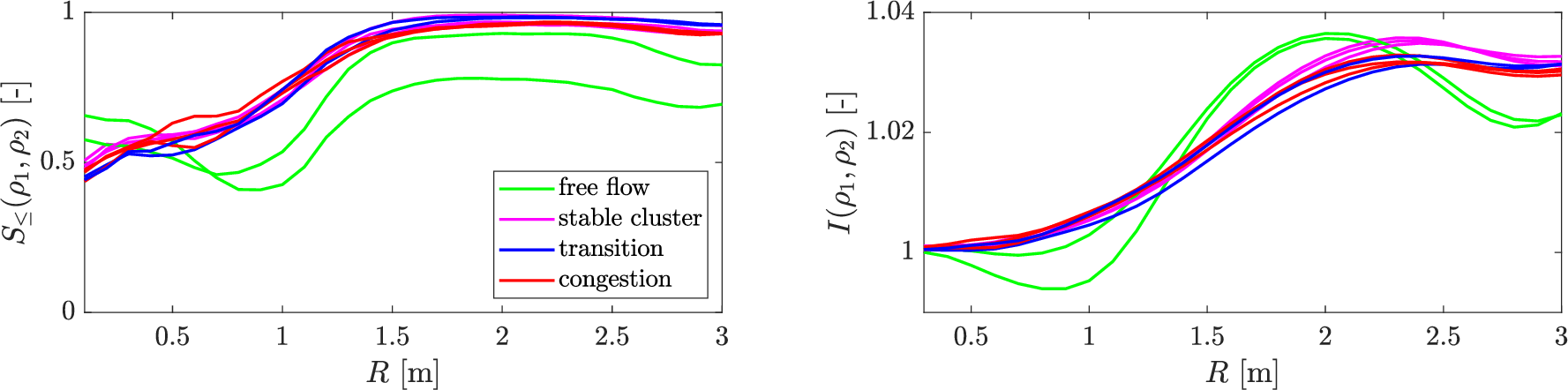}
	\end{center}
	\caption{Success and integral ratios for the conic $\rho_1$ and Borsalino density $\rho_2$ for all experimental rounds (free flow - green, transition - blue, stable cluster - magenta, congestion - red).} \label{fig:borsalino_cone_ratios}
\end{figure}

To be complete, the success (\ref{eq:success_ratio}) and integral ratios (\ref{eq:integral_ratio}) were also computed, which are depicted in Figure \ref{fig:borsalino_cone_ratios}. The success ratio of the Borsalino density is greater than that of the conic density for increasing values of $R$ (for most time steps). Moreover, the integral ratios of the Borsalino and conic kernels belong to the interval $\langle 0.99, 1.04)$ for all $R$ and all experimental rounds, thereby confirming the strong similarity between the densities generated using these two kernels.

We can conclude in this section that the conic and the Borsalino kernels generate almost identical densities. However, the Borsalino kernel can be advantageously used for theoretical studies due to its smoothness (the kernel function is infinitely differentiable).

\subsection{Comparison of Conic and Cylindrical Kernels}
\begin{figure}[h!]
	\begin{center}
	\includegraphics[width=0.55\textwidth]{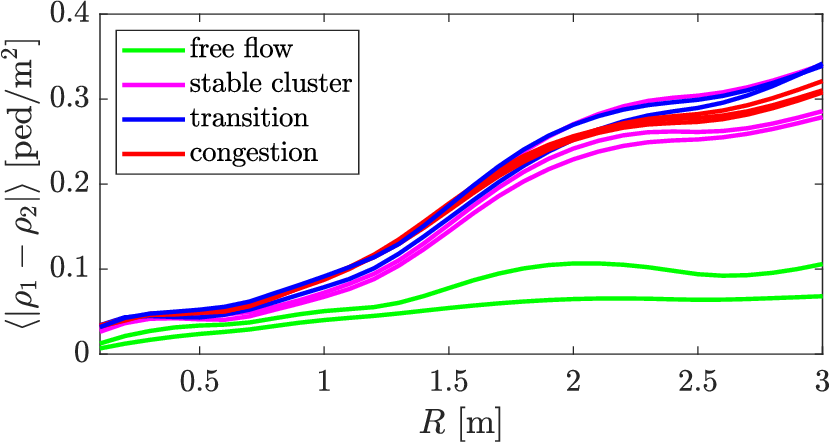}
	\end{center}
	\caption{Mean absolute deviation between the cylindrical $\rho_1$ and conic density $\rho_2$ for all experimental rounds (free flow - green, transition - blue, stable cluster - magenta, congestion - red).} \label{fig:cylinder_cone_meandifference}
\end{figure}

Similarly, we examined the mean absolute deviation (\ref{eq:mean_abs_diff}) between the densities generated by the conic and cylindrical kernels, and the results are shown in Figure \ref{fig:cylinder_cone_meandifference} for all experimental rounds. Cone and cylinder produce similar densities for any blur $R$ for free flow experimental rounds and for $R < 1$ m for other movement phases, since their mean absolute deviation is less than $0.1$ ped/m$^2$. The highest deviation occurs less than $0.4$ ped/m$^2$.

\begin{figure}[h!]
	\begin{center}
	\includegraphics[width=1\textwidth]{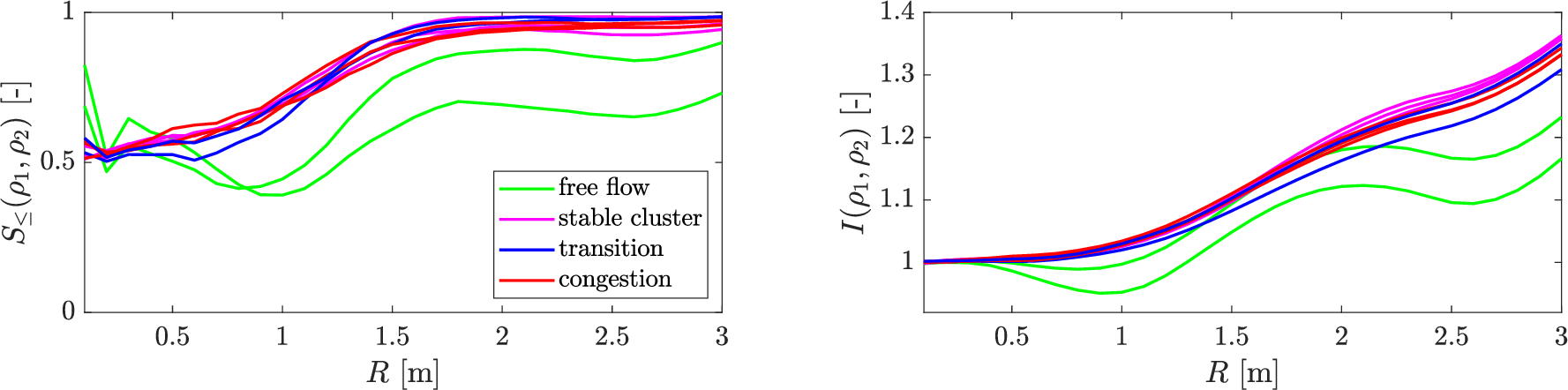}
	\end{center}
	\caption{Success and integral ratios for the cylindrical $\rho_1$ and conic density $\rho_2$ for all experimental rounds (free flow - green, transition - blue, stable cluster - magenta, congestion - red).} \label{fig:cylinder_cone_ratios}
\end{figure}

Additional measures, success (\ref{eq:success_ratio}) and integral ratios (\ref{eq:integral_ratio}), for conic and cylindrical kernels are depicted in Figure \ref{fig:cylinder_cone_ratios}. According to the results, the conic kernel generates a greater density than the cylindrical kernel, particularly for greater values of blur $R$ and higher occupancy in the detector. This behaviour is caused by the linear trend of the cone and the constant trend of the cylinder.

To summarize, the conic kernel coincides with the cylindrical kernel for lower values of blur. On the contrary, the conic density is greater than the cylindrical density for $R \geq 1$ m.

Note that supplementary quantitative measures complementing the discussion about their similarity, including their time development, will be considered in the next section.

\subsection{Comparison of Conic Kernel and Minimum Distance Estimate} \label{sec:mindist_cone}
Firstly, we have to calibrate the minimum distance distributions. The calibration process is separated according to the flow type and the calibration constant may depend on the size of blur $R$. We have to ensure that the training and testing set contain every type of flow, by defining the set as follows: the training set includes round numbers 5, 4, 9, and 7 (free flow, transition, stable cluster, and congestion), and the testing set includes round numbers 2, 6, 10, and 11 (in the same order as the training set).

\begin{figure}[h!]
	\begin{center}
	\includegraphics[width=0.55\textwidth]{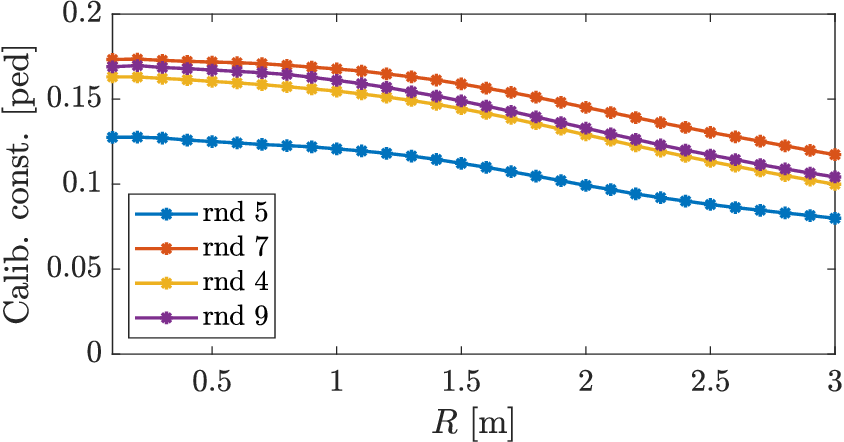}
	\end{center}
	\caption{Calibration constant of minimum distance $1/d^2$ density for different experimental rounds (training data).} \label{fig:mindistcalib}
\end{figure}

\begin{figure}[h!]
	\begin{center}
	\includegraphics[width=1.1\textwidth]{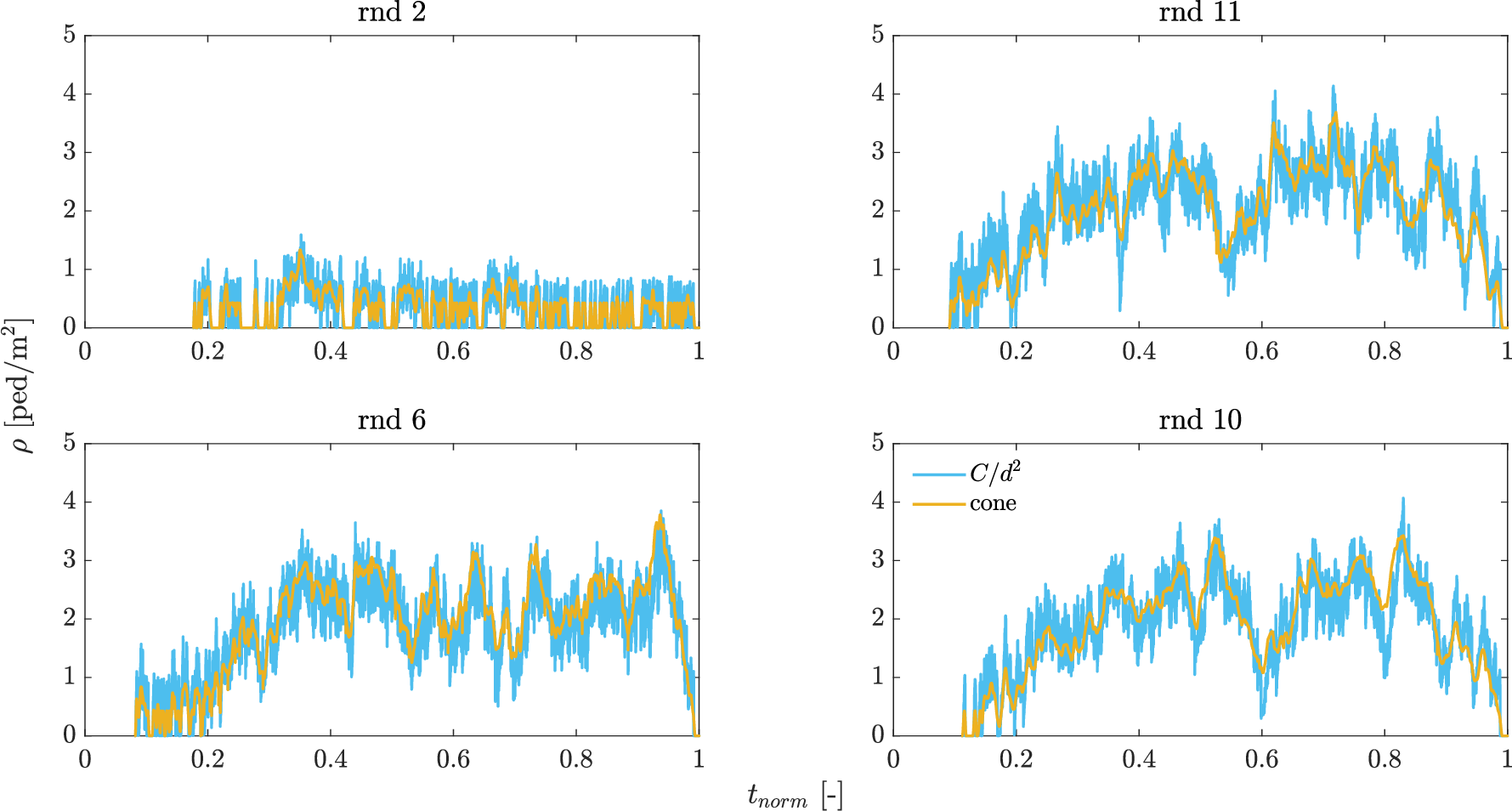}
	\end{center}
	\caption{Comparison of the calibrated minimum distance $1/d^2$ density with the conic density ($R = 0.9$ m) for different experimental rounds using the testing data, i.e., data not used for calibration.} \label{fig:mindist}
\end{figure}

The calibration metric is chosen as the mean absolute deviation (\ref{eq:mean_abs_diff}) between the conic density $\rho_{\textup{cone}}$ and the density of minimum distance distribution $\rho_{\textup{md}}$ in the detector. Thus, the calibration constant for fixed blur $R$ is found as
\begin{equation}
\textup{argmin}_{c_{md} \in \R^+} \langle | \rho_{\textup{cone}}(R) - \rho_{\textup{md}}(c_{md}) | \rangle .
\end{equation}
The calibration constant for minimum distance distribution $1/d^2$ with respect to $R$ is depicted in Figure \ref{fig:mindistcalib}. To avoid the dependency on blur, we approximate the function of the calibration constant with respect to blur $R$ by a constant function with the value belonging to $R = 0.9$ m which needs to be made separately for free flow and other conditions. Then, we obtained the calibration constants for minimum distance distribution $1/d$ as
\begin{equation}
c^{(1)}_{md} = \left\{ \begin{array}{lcl}
0.0233 \, \textnormal{ped} & \dots & \textnormal{free flow}, \\
0.0400 \, \textnormal{ped} & \dots & \textnormal{otherwise},
\end{array} \right.
\end{equation}
and for minimum distance distribution $1/d^2$ as
\begin{equation}
c^{(2)}_{md} = \left\{ \begin{array}{lcl}
0.1219 \, \textnormal{ped} & \dots & \textnormal{free flow}, \\
0.1628 \, \textnormal{ped} & \dots & \textnormal{otherwise}.
\end{array} \right.
\end{equation}
The fit of the calibrated distribution $1/d^2$ on the testing set (using the conic kernel with a fixed $R$) is shown in Figure \ref{fig:mindist}. The correspondence between the curves is evident for every experimental round, representing different phases of the system.

We note that this calibration process using the conic kernel is robust. If any kernel distribution is not available, the calibration process would be performed using the Dirac distribution. However, our process contains the Dirac distribution implicitly, as discussed in Section \ref{sec:converg_dirac}.

\section{Results and Discussion}
Previous bilateral studies have shown that non-parametric density evaluation methods (point approximation, Voronoi diagram, minimum distance estimate) produce similar results to the kernel method with the appropriate value of blur parameter. Specifically, the blur tending to zero corresponds to the point approximation, and when the blur falls within the interval $(1.5, 3)$ m, it fits the Voronoi method. The kernel density corresponds to the density evaluated using the minimum distance to a pedestrian. As theoretically derived, the minimum distance squared in two-dimensional space is more accurate to apply than the one-dimensional formula coming from vehicular traffic.

\begin{figure}[h!]
	\begin{center}
	\includegraphics[width=1.1\textwidth]{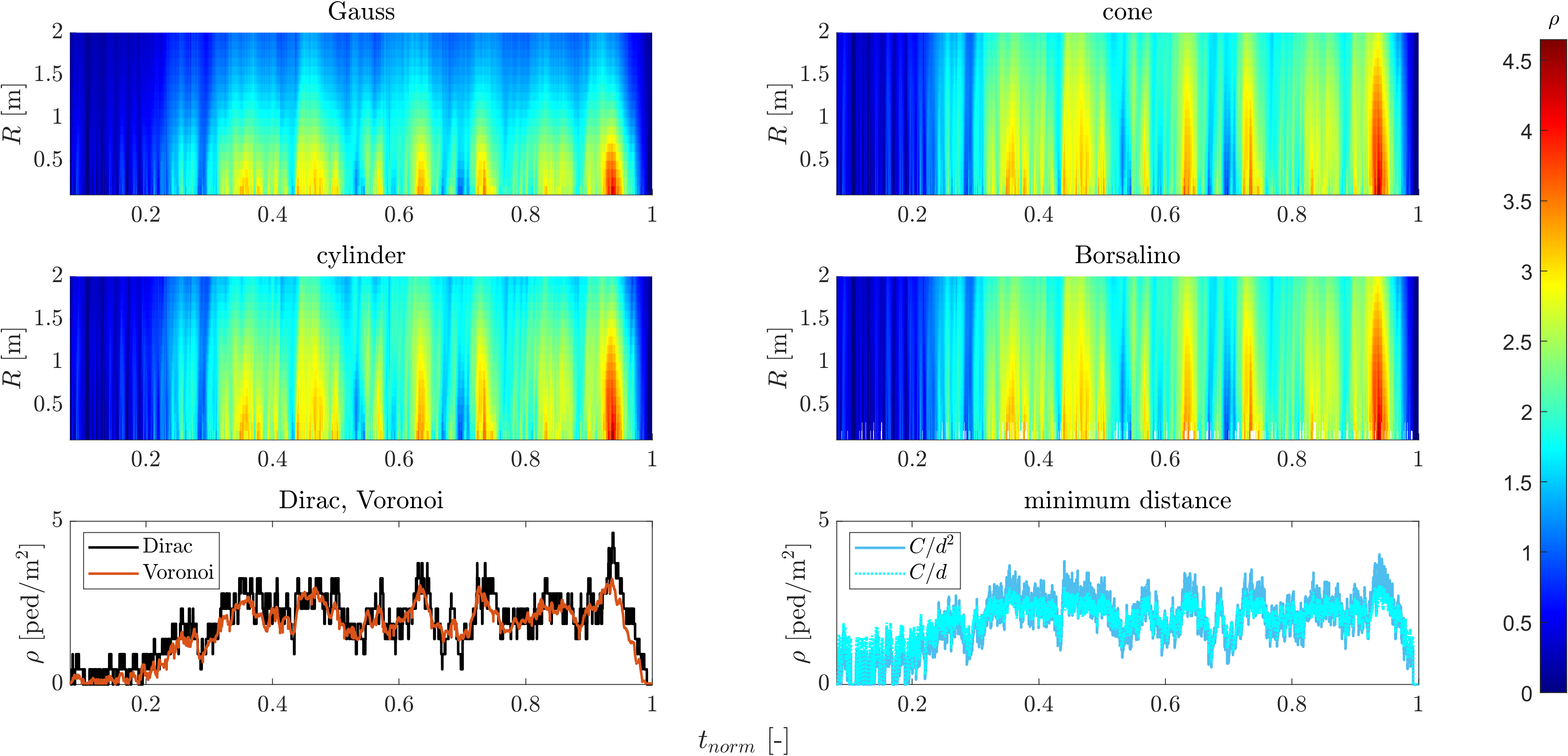}
	\caption{Density with respect to time, kernel and its size (where relevant).} \label{fig:rnd6_count_t_R}
	\end{center}
\end{figure}

The similarity of all approaches is evident from the time development perspective. The pedestrian densities $\rho(t)$ for the examined methods with $R \in \langle 0.1, 2 \rangle$ m (where relevant) evaluated for round 6 of the experiment are plotted in Figure \ref{fig:rnd6_count_t_R}. Density peaks and drops correspond among different methods. The essential difference can be observed between the Gaussian kernel and the conic, cylindrical and Borsalino kernels, which behave similarly. As explained in Section \ref{sec:mass_gauss}, the different scaling causes the Gaussian kernel to significantly smooth the pedestrian density. Non-parametric methods follow the same trend, and the peaks of minimum distance and Voronoi diagram correspond with the kernel estimates.

The results for the minimum distance method are fully comparable with kernel methods in terms of trend perspective and summarized statistics. The variant using the square of the minimum distance ensures better correspondence than the first power of the minimum distance. However, further research is needed to analyse this method in detail and improve the handling of singularity.

\subsection{Roughness, Featurelessness and Stability in Time} \label{sec:rough_feature}
\begin{figure}[h!]
	\begin{center}
	\includegraphics[width=1\textwidth]{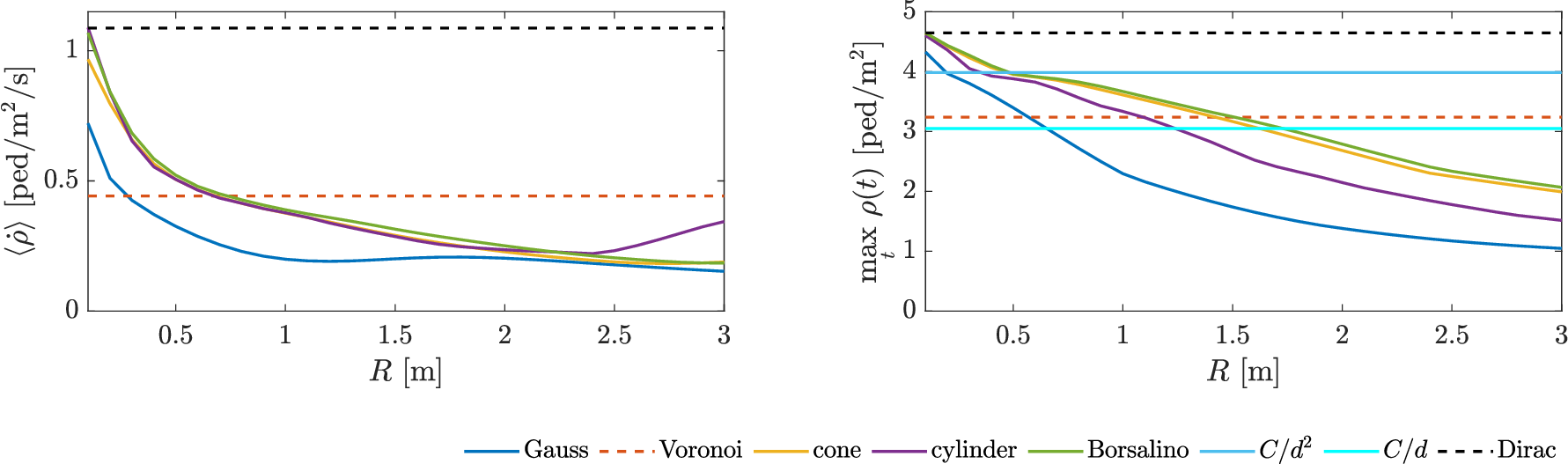}
	\caption{Left: roughness (mean absolute derivative of density). Right: featurelessness (maximum density).} \label{fig:rnd6_mean_der_max_count}
	\end{center}
\end{figure}

Important density properties defined as roughness (\ref{eq:def_roughness}) and featurelessness (\ref{eq:def_featurelessness}) are presented in Figure \ref{fig:rnd6_mean_der_max_count} (for round 6 of the experiment), while a detailed numerical summary can be found in Table \ref{tab:summary_meanabsder_kernel_density_static_detector} and Table \ref{tab:summary_max_kernel_density_static_detector} for different experimental rounds. The properties of minimum distance estimates are depicted only in Figure \ref{fig:rnd6_mean_der_max_count} (right), while their roughness can be found in Table \ref{tab:summary_meanabsder_kernel_density_static_detector} due to their high values outside of the scale of the figure.

The point approximation generates, as the most scattered curve with lots of jumps, an upper boundary for the other methods, except the roughness of the minimum distance estimates, see Figure \ref{fig:rnd6_mean_der_max_count} (left) and Table \ref{tab:summary_meanabsder_kernel_density_static_detector}. It results from the fact that minimum distance distributions have very high peaks near real pedestrian locations. These peaks are similar to the peaks of the density distribution of Dirac kernels. However, the size of these peaks varies in time because they are strongly affected by current pedestrian locations. Furthermore, the Dirac kernel generates a constant section after every jump because the density in the detector does not change for several time steps before another pedestrian leaves or enters the detector. These deviations are cumulated in the roughness of the density, and thus the minimum distance estimate reaches a much higher value than the point approximation.

Voronoi density can be interpreted as the median value for conic, Borsalino or cylindrical curves in terms of roughness, as seen in Figure \ref{fig:rnd6_mean_der_max_count}. Gaussian roughness confirms that the Gaussian kernel is the most smoothed kernel shape without any rescaling. Borsalino and conic curves are in correspondence. The cylindrical kernel has a very similar roughness to the conic kernel, the cylinder is greater than the cone for low values of $R$. This is produced by the fact that the cylinder is more similar to the Dirac distribution than the cone, especially for low values of $R$. The cylindrical kernel converges to the roughness of the conic kernel for increasing $R$, if we exclude the trend from $R > 2.5$ m. This behaviour is caused by the size of the detector and a very smoothed pedestrian leaving this detector (and the rest of their kernel still being in the detector at the same time). Overall, the larger the detector, the smaller the jump in density.

To discuss featurelessness, see Figure \ref{fig:rnd6_mean_der_max_count} (right) and Table \ref{tab:summary_max_kernel_density_static_detector}. The Gaussian kernel is the smoothest kernel, therefore its featurelessness has the lowest value among all compared methods for every blur $R$. Due to the definition of the kernels, the cone produces greater featurelessness than the cylinder for all values of $R$, since the peak of conic density is always higher than that of cylindrical density. For the same reason, Borsalino has greater featurelessness than the cone. Dirac produces the highest value of featurelessness, and Voronoi represents the median featurelessness for conic, cylindrical and Borsalino kernels. For every kernel and every $t \in \R^+$, we have
$$
\max_t \rho_{\textup{Dirac}}(t) \geq \max_t \rho_{\textup{Borsalino}}(t) \geq \max_t \rho_{\textup{Cone}}(t) \geq \max_t \rho_{\textup{Cylinder}}(t) > \max_t \rho_{\textup{Gauss}}(t).
$$
The featurelessness of minimum distance estimates is similar to that of Voronoi in the case of $C/d$ and similar to that of kernel methods in the case of $C/d^2$ for $R < 1$ m.

\begin{table}
\tbl{Summary of the mean absolute derivative of density for the different methods used for the static detector.}
{\begin{tabular}{c|l;{1pt/1pt}c||c|c|c|c||c|c|c|c|} 
\cline{4-11}
\multicolumn{1}{l}{} & \multicolumn{1}{l}{$\langle \dot{\rho} \rangle$ [ped/m$^2$/s]}      & \multicolumn{1}{l|}{} & \multicolumn{4}{c||}{Training data for Min. distance} & \multicolumn{4}{c|}{Testing data for Min. distance}  \\ 
\cline{2-11}
\multicolumn{1}{l|}{} 
& Method 					& R [m] & Rnd 5 & Rnd 7  & Rnd 4  & Rnd 9  & Rnd 2 & Rnd 11 & Rnd 6 & Rnd 10    \\ 
\hline\hline
\multirow{4}{*}{NP}   
& Voronoi  					 & - &  0.38	&	0.35	&	0.43	&	0.34	&	0.17	&	0.37	&	0.44	&	0.36 \\ 
\cline{2-11}
& Point approximation        & - & 1.04	&	1.14	&	1.10	&	1.09	&	0.87	&	1.18	&	1.09	&	1.22 \\ 
\cline{2-11}
& Minimum distance $1/d$     & - & 1.86	&	2.57	&	2.82	&	2.51	&	1.68	&	2.59	&	2.58	&	2.47 \\ 
\cline{2-11}
& Minimum distance $1/d^{2}$ & - & 3.97	&	5.94	&	5.74	&	5.65	&	2.70	&	5.95	&	5.77	&	5.68 \\ 
\hline
\multirow{4}{*}{P1} 
& Conic kernel 				& 0.3 & 0.74	&	0.59	&	0.69	&	0.58	&	0.66	&	0.60	&	0.66	&	0.60 \\ 
\cline{2-11}
& Cylindrical kernel        & 0.3 & 0.70	&	0.57	&	0.68	&	0.58	&	0.61	&	0.59	&	0.65	&	0.57 \\ 
\cline{2-11}
& Borsalino kernel          & 0.3 & 0.77	&	0.61	&	0.72	&	0.61	&	0.68	&	0.62	&	0.68	&	0.62 \\ 
\cline{2-11}
& Gaussian kernel           & 0.1 & 0.80	&	0.66	&	0.75	&	0.65	&	0.70	&	0.67	&	0.72	&	0.67 \\
\hline
\multirow{4}{*}{P2} 
& Conic kernel 			    & 0.9 & 0.43	&	0.29	&	0.41	&	0.30	&	0.38	&	0.29	&	0.39	&	0.30 \\ 
\cline{2-11}
& Cylindrical kernel        & 0.9 & 0.41	&	0.26	&	0.41	&	0.28	&	0.33	&	0.27	&	0.39	&	0.29 \\ 
\cline{2-11}
& Borsalino kernel          & 0.9 & 0.45	&	0.30	&	0.42	&	0.32	&	0.39	&	0.31	&	0.41	&	0.31 \\ 
\cline{2-11}
& Gaussian kernel           & 0.3 & 0.47	&	0.33	&	0.44	&	0.34	&	0.43	&	0.34	&	0.43	&	0.34 \\
\hline
\multirow{4}{*}{P3} 
& Conic kernel 				& 1.5 & 0.31	&	0.18	&	0.30	&	0.20	&	0.23	&	0.19	&	0.29	&	0.20 \\ 
\cline{2-11}
& Cylindrical kernel        & 1.5 & 0.30	&	0.20	&	0.30	&	0.20	&	0.26	&	0.21	&	0.29	&	0.21 \\ 
\cline{2-11}
& Borsalino kernel          & 1.5 & 0.33	&	0.19	&	0.33	&	0.21	&	0.26	&	0.22	&	0.31	&	0.22 \\ 
\cline{2-11}
& Gaussian kernel           & 0.5 & 0.35	&	0.21	&	0.33	&	0.23	&	0.28	&	0.22	&	0.33	&	0.23 \\
\hline
\end{tabular}} \label{tab:summary_meanabsder_kernel_density_static_detector} 
\end{table}

\begin{table}
\tbl{Summary of the maximum density for the different methods used for the static detector.}
{\begin{tabular}{c|l;{1pt/1pt}c||c|c|c|c||c|c|c|c|} 
\cline{4-11}
\multicolumn{1}{l}{} & \multicolumn{1}{l}{Max. density  [ped/m$^2$]}      & \multicolumn{1}{l|}{} & \multicolumn{4}{c||}{Training data for Min. distance} & \multicolumn{4}{c|}{Testing data for Min. distance}  \\ 
\cline{2-11}
\multicolumn{1}{l|}{} 
& Method 					& R [m] & Rnd 5 & Rnd 7  & Rnd 4  & Rnd 9  & Rnd 2 & Rnd 11 & Rnd 6 & Rnd 10    \\ 
\hline\hline
\multirow{4}{*}{NP}   
& Voronoi  					 & - & 1.16	&	3.24	&	2.76	&	2.91	&	0.66	&	3.62	&	3.24	&	3.25 \\ 
\cline{2-11}
& Point approximation        & - & 2.32	&	4.18	&	3.72	&	4.18	&	1.86	&	4.18	&	4.65	&	4.18 \\ 
\cline{2-11}
& Minimum distance $1/d$     & - & 1.53	&	3.09	&	3.00	&	2.97	&	1.19	&	3.10	&	3.05	&	3.15 \\ 
\cline{2-11}
& Minimum distance $1/d^{2}$ & - & 2.16	&	3.96	&	3.87	&	3.89	&	1.58	&	3.96	&	3.99	&	4.03 \\ 
\hline
\multirow{4}{*}{P1} 
& Conic kernel 				& 0.3 & 2.21	&	3.71	&	3.49	&	3.76	&	1.53	&	3.91	&	4.23	&	3.90 \\ 
\cline{2-11}
& Cylindrical kernel        & 0.3 & 2.12	&	3.60	&	3.32	&	3.54	&	1.46	&	3.74	&	4.05	&	3.74 \\ 
\cline{2-11}
& Borsalino kernel          & 0.3 & 2.24	&	3.73	&	3.53	&	3.80	&	1.54	&	3.93	&	4.27	&	3.94 \\ 
\cline{2-11}
& Gaussian kernel           & 0.1 & 2.26	&	3.78	&	3.57	&	3.88	&	1.59	&	4.01	&	4.33	&	4.01 \\
\hline
\multirow{4}{*}{P2} 
& Conic kernel 			    & 0.9 & 1.77	&	3.33	&	3.01	&	3.10	&	1.30	&	3.64	&	3.70	&	3.37 \\ 
\cline{2-11}
& Cylindrical kernel        & 0.9 & 1.65	&	3.21	&	2.89	&	2.91	&	1.20	&	3.49	&	3.43	&	3.25 \\ 
\cline{2-11}
& Borsalino kernel          & 0.9 & 1.81	&	3.37	&	3.03	&	3.14	&	1.33	&	3.68	&	3.75	&	3.39 \\ 
\cline{2-11}
& Gaussian kernel           & 0.3 & 1.87	&	3.42	&	3.05	&	3.20	&	1.34	&	3.68	&	3.80	&	3.47 \\
\hline
\multirow{4}{*}{P3} 
& Conic kernel 				& 1.5  & 1.47	&	3.01	&	2.67	&	2.74	&	1.09	&	3.27	&	3.17	&	3.02 \\ 
\cline{2-11}
& Cylindrical kernel        & 1.5 & 1.22	&	2.73	&	2.35	&	2.50	&	0.92	&	2.91	&	2.67	&	2.75 \\ 
\cline{2-11}
& Borsalino kernel          & 1.5 & 1.53	&	3.07	&	2.73	&	2.78	&	1.12	&	3.34	&	3.25	&	3.09 \\ 
\cline{2-11}
& Gaussian kernel           & 0.5 & 1.57	&	3.13	&	2.82	&	2.88	&	1.17	&	3.43	&	3.40	&	3.17 \\
\hline
\end{tabular}} \label{tab:summary_max_kernel_density_static_detector} 
\end{table}

\begin{table}
\tbl{Summary of the mean density for the different methods used for the static detector.}
{\begin{tabular}{c|l;{1pt/1pt}c||c|c|c|c||c|c|c|c|} 
\cline{4-11}
\multicolumn{1}{l}{} & \multicolumn{1}{l}{Mean density [ped/m$^2$]}  & \multicolumn{1}{l|}{} & \multicolumn{4}{c||}{Training data for Min. distance} & \multicolumn{4}{c|}{Testing data for Min. distance}  \\ 
\cline{2-11}
\multicolumn{1}{l|}{} 
& Method 					& R [m] & Rnd 5 & Rnd 7  & Rnd 4  & Rnd 9  & Rnd 2 & Rnd 11 & Rnd 6 & Rnd 10    \\ 
\hline\hline
\multirow{4}{*}{NP}   
& Voronoi  					 & - &  0.36	&	1.87	&	1.31	&	1.58	&	0.13	&	1.71	&	1.69	&	1.66 \\ 
\cline{2-11}
& Point approximation        & - &  0.70	&	1.99	&	1.55	&	1.77	&	0.32	&	1.94	&	1.88	&	1.86 \\ 
\cline{2-11}
& Minimum distance $1/d$     & - &  0.70	&	1.98	&	1.73	&	1.84	&	0.39	&	1.96	&	1.91	&	1.91 \\ 
\cline{2-11}
& Minimum distance $1/d^{2}$ & - &  0.68	&	1.96	&	1.62	&	1.78	&	0.34	&	1.91	&	1.88	&	1.86 \\ 
\hline
\multirow{4}{*}{P1} 
& Conic kernel 				& 0.3 & 0.74	&	2.09	&	1.62	&	1.85	&	0.34	&	2.02	&	1.98	&	1.96 \\ 
\cline{2-11}
& Cylindrical kernel        & 0.3 & 0.74	&	2.08	&	1.61	&	1.85	&	0.34	&	2.01	&	1.97	&	1.95 \\ 
\cline{2-11}
& Borsalino kernel          & 0.3 & 0.74	&	2.09	&	1.62	&	1.85	&	0.34	&	2.02	&	1.98	&	1.96 \\ 
\cline{2-11}
& Gaussian kernel           & 0.1 & 0.74	&	2.09	&	1.62	&	1.85	&	0.34	&	2.02	&	1.98	&	1.96 \\
\hline
\multirow{4}{*}{P2} 
& Conic kernel 			    & 0.9 & 0.74	&	2.06	&	1.58	&	1.82	&	0.36	&	1.97	&	1.93	&	1.90 \\ 
\cline{2-11}
& Cylindrical kernel        & 0.9 & 0.74	&	2.03	&	1.55	&	1.78	&	0.37	&	1.93	&	1.89	&	1.86 \\ 
\cline{2-11}
& Borsalino kernel          & 0.9 & 0.74	&	2.06	&	1.58	&	1.82	&	0.35	&	1.98	&	1.93	&	1.91 \\ 
\cline{2-11}
& Gaussian kernel           & 0.3 & 0.73	&	2.06	&	1.59	&	1.83	&	0.35	&	1.98	&	1.94	&	1.92 \\
\hline
\multirow{4}{*}{P3} 
& Conic kernel 				& 1.5  & 0.71	&	1.95	&	1.47	&	1.70	&	0.36	&	1.83	&	1.80	&	1.77 \\ 
\cline{2-11}
& Cylindrical kernel        & 1.5 & 0.64	&	1.80	&	1.34	&	1.54	&	0.35	&	1.66	&	1.64	&	1.61 \\ 
\cline{2-11}
& Borsalino kernel          & 1.5 & 0.72	&	1.98	&	1.50	&	1.73	&	0.37	&	1.86	&	1.84	&	1.80 \\ 
\cline{2-11}
& Gaussian kernel           & 0.5 & 0.72	&	2.00	&	1.52	&	1.75	&	0.36	&	1.89	&	1.86	&	1.83 \\
\hline
\end{tabular}} \label{tab:summary_mean_kernel_density_static_detector} 
\end{table}

\clearpage
We have also examined roughness and featurelessness for other experimental rounds, and although we obtained very similar results and trends as for round 6, there is one point to note. Although Dirac sets the upper boundary for kernel methods for roughness and featurelessness in round 6, it may not hold in general. Under specific conditions, it is possible for other methods to exceed this boundary. For instance, if there are $n \in \N$ pedestrians in the detector with the kernel completely inside the detector and at least one pedestrian outside the detector but close to the detector boundary with the kernel distribution partially enriching the density in the detector with $q/|A|$, where $q \in \Q^+, q < 1$. Then the Dirac density in the detector can be evaluated as $n/|A|$ ped/m$^2$. However, the kernel density in the detector is greater than this value, specifically $(n + q)/|A|$ ped/m$^2$. Nevertheless, this situation is quite rare.

The mean density values for all examined methods are shown in Table \ref{tab:summary_mean_kernel_density_static_detector}. It can be seen that the mean density values, as presented in Section \ref{sec:methodology}, are consistent across all examined methods and for all examined parametric sets.

\subsection{Alternative Smoothing Techniques}
Desired 'smooth' values of measured density can be obtained by either by applying advanced, less scatter estimates (as presented in this article) or by filtering the final time series. Specifically, these two options are:
\begin{enumerate}
	\item \textbf{Spatial smoothing}, which replaces the jump-like pedestrian contribution (Dirac delta function) by a spatial representation of a person. This can be achieved by distance-weighted contributions (e.g., by kernels) or a cell approach.
	\item \textbf{Time smoothing}, which balances the rises and drops of the density values. This can be achieved, for example, by using (weighted) moving averages or any other filtration method conserving the mean value of the observed time series.
\end{enumerate}         

\begin{figure}[h!]
	\begin{center}
	\includegraphics[width=\textwidth]{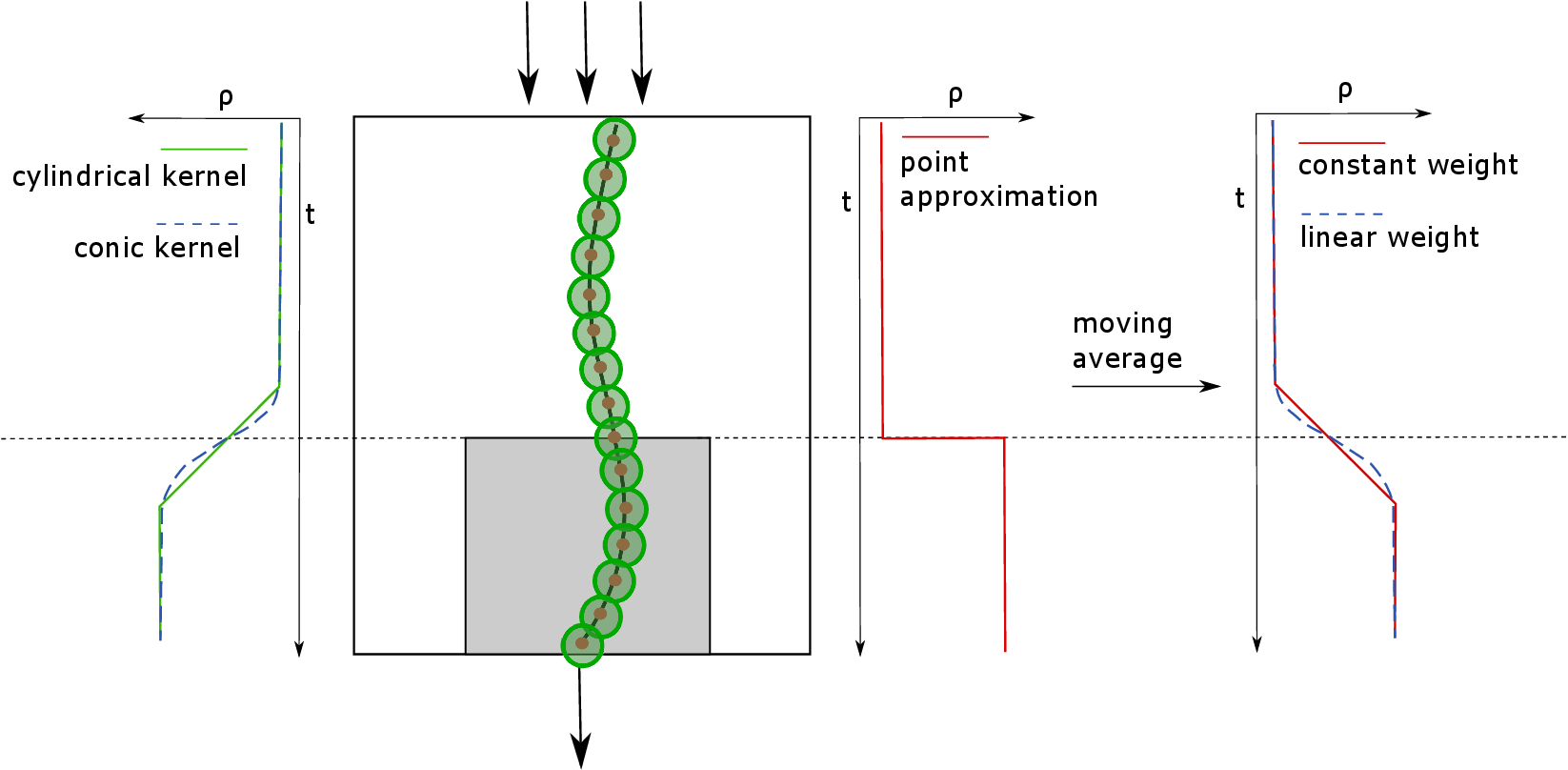}
	\caption{Equivalence of kernel approach (left); point approximation and time smoothing processed by moving average (right). One pedestrian is moving into the detector (grey) area of 1 m$^2$ (middle). The point approximation density in the detector instantly jumps from zero to one as the pedestrian enters. This jump can be smoothed either by space (left) or time (right) averages.}
	\label{fig:smoothing}
	\end{center}
\end{figure}

Weighted moving averages (time smoothing) can be completely equivalent to kernel estimates (spatial smoothing) with respect to the assumption of pedestrian velocity or border conditions. As shown in Figure \ref{fig:smoothing}, the density jumps caused by entering a new pedestrian into the detector can be smoothed either by time or spatial averages, producing the same results with appropriate parametrization.

\begin{figure}[h!]
	\begin{center}
	\includegraphics[width=0.55\textwidth]{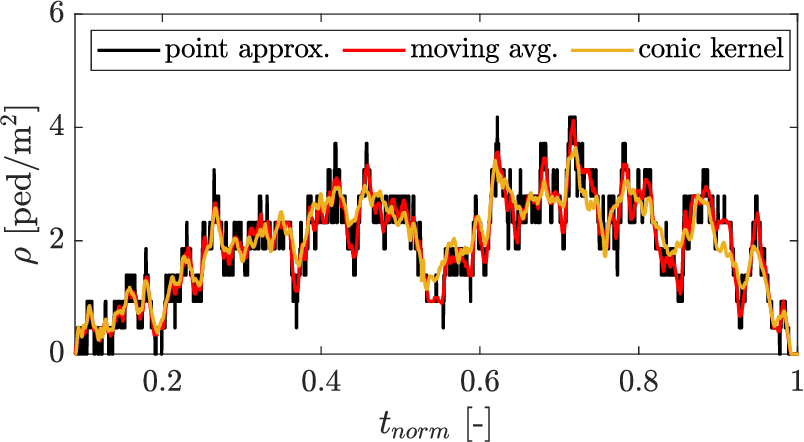}
	\caption{Example of different smoothing techniques: moving averages and kernel distribution.} \label{fig:rnd6_example_smoothing_techniques}
	\end{center}
\end{figure}

The comparison of those techniques applied to experimental data (round 6) is depicted in Figure \ref{fig:rnd6_example_smoothing_techniques}. The figure shows the curve of point approximation (Dirac kernel), conic kernel with $R = 0.9$ m, and moving averages (averaging the values of point approximation belonging to one second). We can see that these smoothing techniques are equivalent and can be calibrated to achieve the best correspondence.

\section{Conclusions}
To conclude, the density evaluated inside a detector significantly depends on the blur parameter $R$ hidden inside an individual density distribution (kernel function). As shown in this study, the desired behaviour of detector density from the perspective of roughness (mean absolute derivative), featurelessness (pedestrian blending), and conservation of mean density (stability in time) may be achieved by a specific setting of this parameter. For instance, a conic kernel with a base radius of $(0.7, 1.2)$ m produces smooth values that still keep trend features. Presented (symmetric) kernels may differ by scaling (e.g., conic kernel versus Gaussian kernel), however, all of them converge (directly or after rescaling) to a similar value of density. The shape itself can be selected based on its essential qualities, implementation preferences, or to fine-tune the density estimate.

The purpose of this paper is not to recommend one method or one specific set of parameters since it is not possible without any specific research context. The specific method and an appropriate value of the kernel parameter should be selected with respect to the desired behaviour expressed by, for instance, roughness and featurelessness. The benefit of the presented general kernel concept using, e.g., a conic kernel is the ability to replicate the results of other methods (minimum distance, Voronoi) by a simple change of one kernel parameter.

The research introduced in this paper presents fundamental results for other work. An individual density distribution was used by an author to quantify pedestrian \emph{comfort} in the civil engineering context in cooperation with the Faculty of Civil Engineering, Brno University of Technology. The definition of comfort is based on density distributions in the way of median value over time. To find the appropriate kernel size, proxemics zones were assumed (\cite{hall1959silent}), and 80 percent of the Gaussian mass was chosen to fit into personal space (which was delimited by Hall). The rest of the kernel covers other proxemics zones. This definition gives reasonable findings according to preliminary results. The details of this project can be seen in \cite{tacrzeta_prubezna}.

Finally, further research will extend the presented work by analysing pedestrian surroundings, i.e., the study of the area affecting pedestrian behaviour, assuming the density distribution is already known. The other area under investigation focuses on minimum distance estimates. Recent experiences with the pandemic situation increased the priority of social distancing, thus, studying minimum distance in detail is worthwhile. Furthermore, kernels with dynamically changing parameters can be studied in the future.

\section*{Acknowledgement}
This work was supported by the Ministry of Education, Youth, and Sports of the Czech Republic (M\v SMT \v CR) under Grant SGS21/165/OHK4/3T/14 and under project Center for Advanced Applied Science CZ.02.1.01/0.0/0.0/16-019/0000778.

\section*{Disclosure statement}
The authors report there are no competing interests to declare.

\bibliographystyle{tfcad}
\bibliography{paper_JVMB}

\end{document}